\documentclass[letterpaper,12pt]{article}   

\usepackage{aop}
\usepackage{bbm}

\usepackage[colorlinks=true,bookmarks=false,citecolor=red,
urlcolor=red,pdfstartview=FitH]{hyperref}

\begin{document}

\title{Quantum state discrimination}

\author{Stephen M. Barnett$^{1,*}$ and  Sarah Croke$^2$}
\address{$^1$Department of Physics, University of Strathclyde \\ Glasgow G4 0NG, UK}
\address{$^2$Perimeter Institute for Theoretical Physics \\ Waterloo, Ontario, N2L 2Y5, Canada}

\address{$^*$Corresponding author: steve@phys.strath.ac.uk}

\newcommand{\bra}[1]{\langle #1|}
\newcommand{\ket}[1]{|#1\rangle}
\newcommand{\braket}[2]{\langle #1|#2\rangle}

\begin{abstract}It is a fundamental consequence of the superposition principle for quantum states that
there must exist non-orthogonal states, that is states that, although different, have a non-zero 
overlap.  This finite overlap means that there is no way of determining with certainty in which of two 
such states a given physical system has been prepared.  We review the various strategies that have 
been devised to discriminate optimally between non-orthogonal states and some of the optical
experiments that have been performed to realise these.
\end{abstract}


\section{Introduction}
The state of a quantum system is a mysterious object and has been the subject of much 
attention since the earliest days of quantum theory.  We know that it provides a way of 
calculating the observed statistical properties of any desired observable but that it is
not, itself observable.  This means that we cannot determine by observation the state 
of any single physical system.  If we have some prior information, however, then we 
may be able to use this to determine, at least to some extent, the state.  Consider, for 
example, a single photon which we know has been prepared with either horizontal or
vertical polarisation.  A suitably oriented polarising beam splitter can be used to transmit
the photon if it was vertically polarised and reflect it if its polarisation was horizontal.  
Determining the path of the photon by absorbing it with a suitable detector then determines
the state to have been one of horizontal or vertical polarisation.

Suppose, however, that we are told that our photon was prepared with either horizontal or
with left-circular polarisation.  These quantum states of polarisation are not orthogonal in
that states of circular polarisation are superpositions of those of both vertical and horizontal
polarisation:
\begin{eqnarray}
\label{EqInt.1}
|L \rangle &=& \frac{1}{\sqrt 2}(|H\rangle + i|V\rangle)  \nonumber \\
\Rightarrow \langle H|L\rangle &=& \frac{1}{\sqrt 2} \:\: \neq 0 .
\end{eqnarray}
If we subject our photon to the polarisation measurement outlined in the preceding 
paragraph then a left-circularly polarised photon will appear to be horizontally polarised
with probability $\frac{1}{2}$ and vertically polarised with the same probability.  

The problem
of discriminating between such states is fundamental to the quantum theory of communications
\cite{Helstrom76,Holevo82,Holevo00,Barnett09} and underlies the secrecy of the now well-reviewed science of quantum cryptography \cite{Phoenix95,Lo,Bouwmeester,Gisin02,Loepp06}.  Indeed, we
can use the connection between quantum state discrimination and quantum communications to 
motivate the problem of state discrimination.  We suppose that two parties, conventionally named
Alice and Bob, wish to communicate using a quantum channel.  To do this Alice selects from a
given set of states, $|\psi_i\rangle$ (or more generally mixed states with density operators $\hat\rho_i$)
with a given set of probabilities $p_i$.  The selected state is encoded in the preparation of 
a given physical system, such as photon polarisation, and this is sent to Bob.  Bob will know both
the set of possible states and the associated preparation probabilities.  His task is to determine,
by means of a suitable measurement, the state selected by Alice and hence the intended 
message.  This, then is the quantum state discrimination problem: how can we best discriminate
among a known set of possible states $|\psi_i\rangle$, each having been prepared with a 
known probability $p_i$.

The quantum state discrimination problem, as posed here, has been the subject of active
theoretical investigation for a long time \cite{Helstrom76,Holevo82,Holevo00,Helstrom67,Helstrom68,Holevo73,Yuen75,Davies78}, 
but is only comparatively recently that experiments have been performed and most of these have been based on optics.  There exist in the literature a number of reviews of and introductions to quantum
state discrimination \cite{Barnett09,Barnett97,Chefles00,Barnett01a,Barnett04,Paris04,Bergou07}.  
Our purpose
in preparing this review is twofold: first to bring the rapidly developing field up to date and
secondly to introduce the idea of state discrimination to a wider audience in optics.  It seems
especially appropriate to do this as it is in simple optical experiments that the ideas are most 
transparent and where most of the important practical developments have been made.

\section{Generalised measurements}

Most of us are introduced to the idea of measurements in quantum theory in a manner
that is, essentially, that formulated by von Neumann \cite{vonNeumann}.  Each observable
property $O$ is associated with a Hermitian operator $\hat O$ (or more precisely a self-adjoint
one) the eigenvalues of which are the possible results of a measurement of $O$.  If the 
eigenvalues and eigenvectors are $o_m$ and $|o_m\rangle$ then we can write the operator
$\hat O$ in the diagonal form
\begin{equation}
\label{EqGen.1}
\hat O = \sum_m o_m |o_m\rangle \langle o_m| .
\end{equation}
If the system to be measured has been prepared in the state $|\psi\rangle$ then the probability
that a measurement of $O$ will give the result $o_m$ is
\begin{equation}
\label{EqGen.2}
P(o_m) = |\langle o_m|\psi\rangle|^2 .
\end{equation}
Consider, for example, a measurement to determine whether the polarisation of a single 
photon is horizontal or vertical.  A suitable operator, corresponding to this measurement, 
would be 
\begin{equation}
\label{EqGen.3}
\hat{Pol} = H|H\rangle\langle H| + V|V\rangle\langle V| .
\end{equation}
The probability that a measurement of this property on a photon prepared in the 
circularly polarised state $|L\rangle$ will give the result $H$, corresponding to 
horizontal polarisation, is
\begin{equation}
\label{EqGen.4}
P(H) = |\langle H|L\rangle|^2 = \frac{1}{2} .
\end{equation}

It is helpful, in what follows, to rewrite the above probabilities as the expectation value
of an operator.  In this way the probability that a measurement of optical polarisation
shows the photon to be horizontally polarised is
\begin{equation}
\label{EqGen.5}
P(H) = \left\langle |H\rangle \langle H|\right\rangle = \langle \hat P_H \rangle,
\end{equation}
where $\hat P_H = |H\rangle \langle H|$, the projector onto the state $|H\rangle$,
is the required operator.  More generally, for our operator $\hat O$, the probability
that a measurement gives the value $o_m$ is
\begin{equation}
\label{EqGen.6}
P(o_m) = \left\langle |o_m\rangle \langle o_m|\right\rangle = \langle \hat P_m\rangle .
\end{equation}
We note that the measured value, itself, makes no explicit appearance in this probability;
it is not the eigenvalue but only the corresponding eigenvector that determines the form
of the projector and hence the probability for the associated measurement outcome.  

The projectors have four important mathematical properties and it is helpful to list these:

\begin{itemize}
\item The projectors are Hermitian operators, $\hat P^\dagger_m = \hat P_m $.
This property is associated with the fact that probabilities are, themselves, observable 
quantities.

\item They are positive operators, which means that $\langle \psi| \hat P_m |\psi\rangle 
\ge 0$ for all possible states $|\psi\rangle$.  This reflects the fact that the expectation
value of the projector is a probability and must, therefore, be positive or zero.

\item They are complete in that $\sum_m  \hat P_m = \hat{\mathbbmss{1}}$, so that the sum of the 
probabilities for all possible measurement outcomes is unity.

\item They are orthonormal in that $ \hat P_m \hat P_n = 0$ unless $m=n$.  This 
property is sometimes associated with the fact that measurement outcomes must
be distinct (you can only get one of them).  This view is, as we shall see, not correct.
You can indeed only get one outcome but this does not require the orthonormality 
property.

\end{itemize}

The theory of generalised measurements can be formulated simply by dropping the
final requirement.  To see how this works, we introduce a set of probability operators
$\{\hat \pi_m\}$, each of which we wish to associate with a measurement outcome such
that the probability that our measurement gives the result labeled $m$ is
\begin{equation}
\label{EqGen.7}
P(m) = \langle \hat \pi_m\rangle .
\end{equation}
We insist on the first three of the properties of the projectors, as these are required
if we are to maintain the probability interpretation (\ref{EqGen.7}), but drop the 
final requirement so that our probability operators have the properties:

\begin{itemize}
\item The probability operators are Hermitian: $\hat \pi^\dagger_m = \hat \pi_m $.

\item They are positive operators: $\langle \psi| \hat \pi_m |\psi\rangle 
\ge 0$ for all possible states $|\psi\rangle$.  

\item They are complete: $\sum_m  \hat \pi_m = \hat{\mathbbmss{1}}$.

\end{itemize}

The set of probability operators characterising the possible outcomes of any generalised 
measurement is called a probability operator measure, usually abbreviated to POM
\cite{Helstrom76,Barnett09}.  You will often find this set referred to as a positive operator-valued
measure or POVM \cite{Peres93,Busch95}.  If the latter name is used then the
probability operators become elements of a POVM.  

The differences between the projectors and
more general probability operators are best appreciated by reference to some simple examples
and these will be given in the following sections.  There are, however, some important and
perhaps even surprising points and it is sensible to emphasise these here.  Firstly, the three
properties described above have a remarkable generality in that (i) any measurement can 
be described by the appropriate set of probability operators and (ii) any set of operators that
satisfy the three properties correspond to a possible measurement \cite{Barnett09,Peres93}.
This means that we can seek the optimum measurement in any given situation mathematically,
by searching among all sets of operators that satisfy the required properties.  Having found this
we know that a physical realisation of this will exist and can seek a way to implement it in the
laboratory.  The second point to emphasise is that the number of (orthogonal) projectors can
only be less than or equal to the dimension of the state space.  For optical polarisation, for
example, there are only two orthogonal polarisations and the state space is therefore two-dimensional.
It follows that any von Neumann measurement of polarisation can only have two outcomes.
By dropping the requirement for orthogonality, we allow a generalised measurement to have
any number of outcomes, so a generalised measurement of polarisation can have three, four or 
more different outcomes.  Finally, a generalised measurement allows us to describe the 
simultaneous observation of incompatible observables, such as position and momentum or,
in the context of quantum optics, orthogonal field quadratures \cite{She66,Stenholm92}.
Perhaps the first reported generalised optical measurement was of precisely this form
\cite{Walker84,Walker87}.

\section{State Discrimination -Theory}
\subsection{Mimimum Error Discrimination}
\begin{figure}[h]
\begin{center}
\includegraphics[width=80truemm]{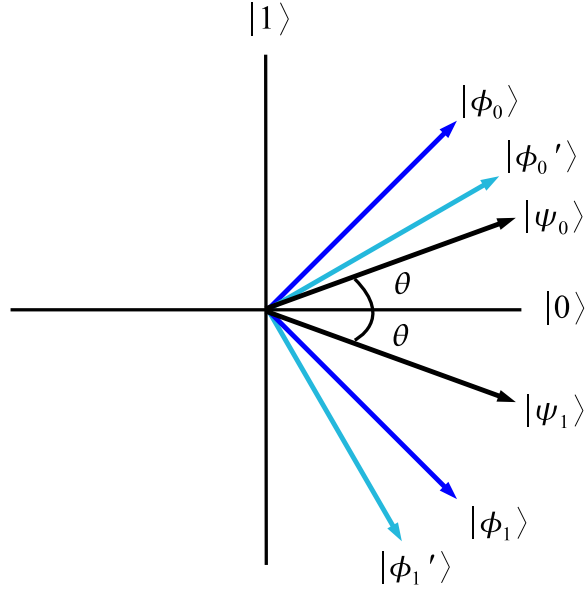}
\end{center}
\caption{The optimal minimum error measurement for discriminating between the pure states $\ket{\psi_0}$, $\ket{\psi_1}$ is a von Neumann measurement.  For $p_0=p_1=1/2$ this is a projective measurement onto the states $\ket{\phi_0}$, $\ket{\phi_1}$, symmetrically located either side of the signal states, and shown in blue here.  For $p_0 > p_1$ the optimal measurement performs better when state $\ket{\psi_0}$ is sent, shown here in light blue (labeled $\ket{\phi_0^{\prime}}$, $\ket{\phi_1^{\prime}}$) is the case $p_0 = 3/4$.}
\label{minerr}
\end{figure}
In quantum state discrimination we wish to design a measurement to distinguish optimally between a given set of states.  As we have seen in the previous section, any physically realisable measurement can be described by a probability operator measure.  Thus by mathematically formulating a figure of merit describing the performance of a measurement, we can search for the set of probability operators describing the optimal measurement.  There are several possible figures of merit, each one corresponding to a different strategy.  Possibly the simplest criteria which may be applied is to minimise the probability of making an error in identifying the state.  We begin with the special case where the state is known to be one of two possible pure states, $\ket{\psi_0}$, $\ket{\psi_1}$, with associated probabilities $p_0$, $p_1=1-p_0$.  If outcome $0$, associated with the probability operator $\hat{\pi}_0$ is taken to indicate that the state was $\ket{\psi_0}$, and outcome $1$ (associated with $\hat{\pi}_1$) is taken to indicate that the state was $\ket{\psi_1}$, the probability of making an error in determining the state is given by
\begin{eqnarray}
P_{err} &=& P(\psi_0) {\rm P}(1|\psi_0) + P(\psi_1) {\rm P}(0|\psi_1) \nonumber \\
&=& p_0 \bra{\psi_0} \hat{\pi}_1 \ket{\psi_0} + p_1 \bra{\psi_1} \hat{\pi}_0 \ket{\psi_1} \nonumber \\
&=& p_0 - {\rm Tr} \left((p_0 \ket{\psi_0} \bra{\psi_0} - p_1 \ket{\psi_1} \bra{\psi_1}) \hat{\pi}_0 \right),
\end{eqnarray}
where in the last line we have used the completeness condition $\hat{\pi}_0 + \hat{\pi}_1 = \hat{\mathbbmss{1}}$.
This expression takes its minimum value when the second term reaches a maximum, which in turn is achieved if $\hat{\pi}_0$ is a projector onto the positive eigenket of the operator $p_0 \ket{\psi_0} \bra{\psi_0} - p_1 \ket{\psi_1} \bra{\psi_1}$.  Note that two pure states define a two-dimensional space, and without loss of generality we can choose an orthogonal basis $\{ \ket{0}, \ket{1} \}$ of this space such that the components of each state in this basis are real.  Thus we can express $\ket{\psi_0}$, $\ket{\psi_1}$ as follows
\begin{eqnarray}
\label{Twostates}
\ket{\psi_0} &=& \cos \theta \ket{0} + \sin \theta \ket{1} \nonumber \\
\ket{\psi_1} &=& \cos \theta \ket{0} - \sin \theta \ket{1},
\label{twopure}
\end{eqnarray}
and the eigenvalues of $p_0 \ket{\psi_0} \bra{\psi_0} - p_1 \ket{\psi_1} \bra{\psi_1}$ can be calculated directly as
\begin{equation}
\lambda_{\pm} = \frac{1}{2} \left( p_0 - p_1 \pm \sqrt{1 - 4 p_0 p_1 \cos^2 2 \theta} \right).
\end{equation}
The minimum probability of making an error is then given by the so-called \textbf{Helstrom bound} \cite{Helstrom76}
\begin{equation}
\label{PHelstrom}
P_{err} = \frac{1}{2}\left( 1 - \sqrt{1 - 4 p_0 p_1 |\braket{\psi_0}{\psi_1}|^2} \right),
\end{equation}
and the optimal measurement is simply a von Neumann measurement.  In particular, for $p_0 = p_1 = 1/2$, the optimal measurement is a projective measurement onto the states
\begin{eqnarray}
\label{Twomeas}
\ket{\phi_0} &=& \frac{1}{\sqrt{2}} \left( \ket{0} + \ket{1} \right) \nonumber \\
\ket{\phi_1} &=& \frac{1}{\sqrt{2}} \left( \ket{0} - \ket{1} \right).
\end{eqnarray}
These are symmetrically located about the signal states, as may be expected from the symmetry of the problem.  As $p_0$ is increased, $\ket{\phi_0}$ moves closer to $\ket{\psi_0}$, and the optimal measurement becomes biased towards making less errors when the more probable state is sent (see Fig. \ref{minerr}).  Finally, as may be expected intuitively, if $p_0$ is much bigger than $p_1$, the optimal measurement is very close to simply asking ``is the state $\ket{\psi_0}$ or not?"

\subsubsection{The minimum error conditions}
The above analysis is easily extended to two mixed states $\hat{\rho}_0$, $\hat{\rho}_1$, in which case the optimal measurement becomes a projective measurement on to the subspaces corresponding to positive and negative eigenvalues of $p_0 \hat{\rho}_0 - p_1 \hat{\rho}_1$.  In the general case of $N$ possible states $\{ \hat{\rho}_i \}$ with associated \emph{a priori} probabilies $\{ p_i \}$, the aim is to minimise the expression
\begin{equation}
P_{err} = \sum_{i=0}^{N-1} p_i \sum_{j \neq i} {\rm Tr}\left( \hat{\rho}_i \hat{\pi}_j \right),
\end{equation}
or equivalently to maximise
\begin{equation}
P_{corr} = 1 - P_{err} = \sum_{i=0}^{N-1} p_i {\rm Tr}\left( \hat{\rho}_i \hat{\pi}_i \right).
\label{pcorr}
\end{equation}
The optimal measurement is known only in certain special cases, however necessary and sufficient conditions which must be satisfied by the optimal POM for the general case are known \cite{Helstrom76,Holevo73,Yuen75} and are given in equations (\ref{hel1},\ref{hel2}).  For simplicity, we prove only sufficiency of the conditions here, but we note that there is also a straight-forward proof of their necessity \cite{Barnett08}.
\begin{highlight}
Necessary and sufficient conditions which must be satisfied by the POM achieving minimum error in distinguishing between the states $\{ \hat{\rho}_i \}$, occuring with probabilities $\{ p_i \}$ are given by
\begin{eqnarray}
\sum_i p_i \hat{\rho}_i \hat{\pi}_i - p_j \hat{\rho}_j &\geq& 0, \quad \forall \quad j \label{hel1} \\
\hat{\pi}_i \left( p_i \hat{\rho}_i - p_j \hat{\rho}_j \right) \hat{\pi}_j &=& 0, \quad \forall \quad i,j. \label{hel2}
\end{eqnarray}
Note that these conditions are not independent, the second may be derived from the first, as shown in the text.
\end{highlight}
If $\{ \hat{\pi}_i \}$ corresponds to an optimal measurement, then for all other POMs $\{ \hat{\pi}_i^{\prime} \}$ we require
\begin{equation}
\sum_i p_i {\rm Tr} (\hat{\rho}_i \hat{\pi}_i) \geq \sum_j p_j {\rm Tr} (\hat{\rho}_j \hat{\pi}_j^{\prime})
\end{equation}
Inserting the identity $\sum_j \hat{\pi}_j^{\prime} = \hat{\mathbbmss{1}}$ gives
\begin{equation}
\sum_j {\rm Tr} \left(\left(\sum_i p_i \hat{\rho}_i \hat{\pi}_i - p_j \hat{\rho}_j \right) \hat{\pi}_j^{\prime} \right) \geq 0.
\end{equation}
Note that $\hat{\pi}_j^{\prime} \geq 0$, thus the above holds if equation (\ref{hel1}) holds, which is therefore a sufficient condition.

For any POM satisfying this condition, it follows that the operator $\hat{\Gamma} = \sum_i p_i \hat{\rho}_i \hat{\pi}_i$ is positive, and therefore Hermitian.  Thus we have
\begin{equation}
\sum_j \left( \sum_i p_i \hat{\pi}_i \hat{\rho}_i - p_j \hat{\rho}_j \right) \hat{\pi}_j = \sum_i \hat{\pi}_i \left(p_i \hat{\rho}_i - \sum_j p_j \hat{\rho}_j \hat{\pi}_j \right) = 0,
\end{equation}
where we have used the fact that the probability operators form a resolution of the identity $\sum_i \hat{\pi}_i = \hat{\mathbbmss{1}}$.  As both $\sum_i p_i \hat{\pi}_i \hat{\rho}_i - p_j \hat{\rho}_j$ and $\hat{\pi}_j$ are positive operators, each term in the sum over $j$ must be identically zero.  Using similar reasoning we can argue that each term in the sum over $i$ must be identically zero.  Thus, in terms of $\hat{\Gamma}$ we obtain
\begin{equation}
(\hat{\Gamma} - p_j \hat{\rho}_j) \hat{\pi}_j = \hat{\pi}_i (p_i \hat{\rho}_i - \hat{\Gamma}) = 0, \quad \forall \, i,j.
\end{equation}
Eliminating $\hat{\Gamma}$ gives equation (\ref{hel2}), which is therefore also a sufficient condition.

\subsubsection{Square Root Measurement}
For any given set of states we can construct an associated measurement, the square root measurement \cite{Kholevo79,Hughston93,Hausladen94,Hausladen96}, as follows
\begin{equation}
\hat{\pi}_i = p_i \hat{\rho}^{-1/2} \hat{\rho}_i \hat{\rho}^{-1/2}
\end{equation}
where $\hat{\rho} = \sum_i p_i \hat{\rho}_i$.  It is clear that the probability operators $\{ \hat{\pi}_i \}$ are positive and sum to the identity, and thus form a complete measurement.  For many of the cases in which the optimal minimum error measurement is known, it is the square root measurement \cite{Ban97,Sasaki98,Eldar01,Barnett01,Chou03,Eldar04}.  We will present here the example of $N$ symmetric pure states occuring with equal \emph{a priori} probabilities $p_i = \frac{1}{N}$, considered by Ban \emph{et al} \cite{Ban97}, and given by
\begin{equation}
\ket{\psi_i} = \hat{V} \ket{\psi_{i-1}} = \hat{V}^i \ket{\psi_0}, \quad i=0,\ldots,N-1
\end{equation}
for some unitary operator $\hat{V}$ satisfying $\hat{V}^N = \hat{\mathbbmss{1}}$.  For this set we have
\begin{equation}
\hat{\rho} = \frac{1}{N} \sum_{i=0}^{N-1} \ket{\psi_i} \bra{\psi_i} = \frac{1}{N} \sum_{i=0}^{N-1} \hat{V}^i \ket{\psi_0} \bra{\psi_0} \hat{V}^{\dagger \, i}
\end{equation}
and it is useful to note that
\begin{equation}
\begin{array}{ccl}
\hat{V} \hat{\rho} \hat{V}^{\dagger} &=& \frac{1}{N} \sum_{i=0}^{N-1} \hat{V} \ket{\psi_i} \bra{\psi_i} \hat{V}^{\dagger} \\
&=& \frac{1}{N} \sum_{i=0}^{N-1} \hat{V}^{i+1} \ket{\psi_0} \bra{\psi_0} \hat{V}^{\dagger \, i+1} \\
&=& \frac{1}{N} \left(\sum_{i=1}^{N-1} \hat{V}^i \ket{\psi_0} \bra{\psi_0} \hat{V}^{\dagger \, i} + \hat{V}^N \ket{\psi_0} \bra{\psi_0} \hat{V}^{\dagger \, N} \right) \\
&=& \hat{\rho}
\end{array}
\end{equation}
where in the last line we have used the property $\hat{V}^N = \hat{\mathbbmss{1}}$.  Thus
\begin{equation}
\hat{V} \hat{\rho} = \hat{V} \hat{\rho} \hat{V}^{\dagger} \hat{V} = \hat{\rho} \hat{V}
\end{equation}
and $\hat{\rho}$ commutes with $\hat{V}$.  The square root measurement consists of the operators
\begin{equation}
\hat{\pi}_i = \frac{1}{N} \hat{\rho}^{-1/2} \ket{\psi_i} \bra{\psi_i} \hat{\rho}^{-1/2} = \frac{1}{N} \hat{\rho}^{-1/2} \hat{V}^i \ket{\psi_0} \bra{\psi_0} \hat{V}^{\dagger \, i} \hat{\rho}^{-1/2},
\end{equation}
and condition (\ref{hel2}) is equivalent to the requirement
\begin{equation}
\bra{\psi_i} \hat{\rho}^{-1/2} \ket{\psi_i} \bra{\psi_i} \hat{\rho}^{-1/2} \ket{\psi_j} - \bra{\psi_i} \hat{\rho}^{-1/2} \ket{\psi_j} \bra{\psi_j} \hat{\rho}^{-1/2} \ket{\psi_j} = 0.
\end{equation}
Noting that
\begin{equation}
\bra{\psi_i} \hat{\rho}^{-1/2} \ket{\psi_i} = \bra{\psi_0} \hat{V}^i \hat{\rho}^{-1/2} \hat{V}^{\dagger \, i} \ket{\psi_0} = \bra{\psi_0} \hat{\rho}^{-1/2} \ket{\psi_0}, \quad \forall \, i,
\end{equation}
we see that this requirement is satisfied.  We now proceed to evaluate $\hat{\Gamma}$
\begin{equation}
\begin{array}{ccl}
\hat{\Gamma} &=& \frac{1}{N} \sum_{i=0}^{N-1} \ket{\psi_i} \bra{\psi_i} \frac{1}{N} \hat{\rho}^{-1/2} \ket{\psi_i} \bra{\psi_i} \hat{\rho}^{-1/2} \\
&=& \frac{1}{N} \bra{\psi_0} \hat{\rho}^{-1/2} \ket{\psi_0} \sum_{i=0}^{N-1} \frac{1}{N} \ket{\psi_i} \bra{\psi_i} \hat{\rho}^{-1/2} \\
&=& \frac{1}{N} \bra{\psi_0} \hat{\rho}^{-1/2} \ket{\psi_0} \hat{\rho}^{1/2}
\end{array}
\end{equation}
To satisfy condition (\ref{hel1}) we require
\begin{equation}
\bra{\phi} \left( \hat{\Gamma} - \frac{1}{N} \ket{\psi_i} \bra{\psi_i} \right) \ket{\phi} \geq 0, \quad \forall \; i,\ket{\phi}.
\end{equation}
Writing $\hat{\Gamma} = \frac{1}{N} \bra{\psi_i} \hat{\rho}^{-1/2} \ket{\psi_i} \hat{\rho}^{1/2}$ we can show
\begin{equation}
\begin{array}{ccl}
\bra{\phi} \hat{\Gamma} \ket{\phi} &=& \frac{1}{N} \bra{\psi_i} \hat{\rho}^{-1/2} \ket{\psi_i} \bra{\phi} \hat{\rho}^{1/2} \ket{\phi} \\
&=& \frac{1}{N} \bra{\psi_i} \hat{\rho}^{-1/4} \hat{\rho}^{-1/4} \ket{\psi_i} \bra{\phi} \hat{\rho}^{1/4} \hat{\rho}^{1/4} \ket{\phi} \\
&\geq& \frac{1}{N} |\bra{\psi_i} \hat{\rho}^{-1/4} \hat{\rho}^{1/4} \ket{\phi}|^2 \\
&=& \frac{1}{N} |\braket{\psi_i}{\phi}|^2,
\end{array}
\end{equation}
where we have used the Cauchy-Schwarz inequality.  Thus condition (\ref{hel1}) holds, and the square root measurement is optimal.  Note that the case of two equiprobable pure states discussed above is an example of a symmetric set.  In this case $\hat{U} = \hat{\sigma}_z$, and it may easily be verified that $\hat{\sigma}_z \ket{\psi_0} = \ket{\psi_1}$ and $\hat{\sigma}_z^2 = \hat{\mathbbmss{1}}$.  Another example of a symmetric set is the so-called trine ensemble \cite{Peres90,Hausladen94}, given by
\begin{eqnarray}
\label{TrineDef}
\ket{\psi_0} &=& \ket{0}  \nonumber \\
\ket{\psi_1} &=& - \frac{1}{2} \ket{0} + \frac{\sqrt{3}}{2} \ket{1} \\
\ket{\psi_2} &=& - \frac{1}{2} \ket{0} - \frac{\sqrt{3}}{2} \ket{1} \nonumber
\end{eqnarray}
and obtained from one another by rotation through an angle of $\frac{2 \pi}{3}$.  These states form a resolution of the identity, and the square root measurement consists of equally weighted projectors onto the states themselves, $\hat{\pi}_i = \frac{2}{3} \ket{\psi_i} \bra{\psi_i}$.

The above solution has been extended to multiply symmetric states \cite{Barnett01} and mixed states \cite{Chou03,Eldar04}.  The square root measurement has also been generalised by Mochon \cite{Mochon06}, who considered measurements of the form
\begin{equation}
\hat{\pi}_i = \hat{\sigma}^{-1/2} p_i^{\prime} \hat{\rho}_i \hat{\sigma}^{-1/2},
\end{equation}
where $\hat{\sigma} = \sum_i p_i^{\prime} \hat{\rho}_i$, i.e. the square-root measurements corresponding to the same set of states but constructed using different \emph{a priori} probabilities.  For pure states, each such measurement is optimal for at least one discrimination problem with the same states, occuring with probabilities given analytically in \cite{Mochon06}.

\subsubsection{Other Results}
Most of the known results for minimum error discrimination correspond to one of the two cases discussed above: that of just two states, or those for which the square-root measurement is optimal.  Another example which is interesting to note is the no-measurement strategy \cite{Hunter03}.  Sometimes the optimal discrimination strategy is not to measure at all, but just to guess the state which is \emph{a priori} most likely, a measurement which may be described by the POM $\{ \hat{\pi}_i = \hat{\mathbbmss{1}}, \hat{\pi}_j = 0, ~ \forall ~ j \neq i \}$, where $i$ is such that $p_i \geq p_j, \forall ~ j$.  Condition (\ref{hel2}) holds trivially for this POM.  Thus the no measurement solution is optimal when condition (\ref{hel1}) holds, which then reads
\begin{equation}
p_i \hat{\rho}_i - p_j \hat{\rho}_j \geq 0, \quad \forall ~ j.
\label{nomeas}
\end{equation}
Clearly this is never optimal if $\hat{\rho}_i$ is pure; a necessary (but not sufficient) condition is that the eigenvectors of $\hat{\rho}_i$ span the entire Hilbert space in which the states $\{ \hat{\rho}_j \}$ lie.  A practical example is discriminating signal states from random noise, described by the density operator $\hat{\rho}_i \propto \hat{\mathbbmss{1}}$.  If the signal to noise ratio is small enough, the minimum error strategy is to always guess that the state received was random noise \cite{Hunter03}.  It is therefore useful to know the noise threshold at which this occurs, which may be deduced from the condition (\ref{nomeas}).

Other examples for which explicit results are known include three mirror symmetric qubit states, both for pure \cite{Andersson02}, and mixed states \cite{Chou04}, and the case of equi-probable pure states, a weighted sum of which equals the identity operator \cite{Yuen75}.  The form of the solution for any set of qubit states has also been explored in some detail by Hunter \cite{Hunterthesis,Hunter04}, including a complete characterisation of the solution for equiprobable pure qubit states.  In the general case, for which explicit results are not known, it is possible to deduce both upper \cite{Barnum02,Montanaro07}, and lower \cite{Montanaro08,Qiu08} bounds on the error probability.  Alternatively, numerical algorithms exist which can find the optimal measurement for a specified set of states to within any desired accuracy \cite{Jezek02,Eldar03}.

\subsection{Unambiguous Discrimination}
In the minimum error measurement, each possible outcome is taken to indicate some corresponding state.  It is perhaps surprising that it is sometimes advantageous to allow for measurement outcomes which don't lead us to identify any state.  Suppose again that we wish to discriminate between the two pure states given by equation (\ref{twopure}), occuring with \emph{a priori} probabilities $p_0$, $p_1$.  Consider the von Neumann measurement
\begin{eqnarray}
\hat{\pi}_? &=& \ket{\psi_1} \bra{\psi_1} \nonumber \\
\hat{\pi}_0 &=& \left(\sin \theta \ket{0} + \cos \theta \ket{1} \right) \left(\sin \theta \bra{0} + \cos \theta \bra{1} \right).
\label{vNunamb}
\end{eqnarray}
If outcome $?$, associated with the probability operator $\hat{\pi}_?$ is realised, we cannot say for sure what state was prepared.  However, note that $\bra{\psi_1} \hat{\pi}_0 \ket{\psi_1} = 0$, and thus when outcome $0$, corresponding to POM element $\hat{\pi}_0$, is realised, we can say for certain that the state was $\ket{\psi_0}$.  Thus, by allowing for measurement outcome $?$, which does not lead us to identify any state, we can construct a measurement which sometimes allows us to determine unambiguously which state was prepared.  This measurement however only ever identifies state $\ket{\psi_0}$, ideally we would like to design a measurement which can identify either state unambiguously, at the cost of sometimes giving an inconclusive result.  The generalised measurement formalism outlined above allows for exactly such a measurement, a possibility that was first pointed out in the seminal papers of Ivanovic \cite{Ivanovic87}, Dieks \cite{Dieks88}, and Peres \cite{Peres88}.

Consider therefore the operators
\begin{eqnarray}
\hat{\pi}_0 &=& a_0 \left(\sin \theta \ket{0} + \cos \theta \ket{1} \right) \left(\sin \theta \bra{0} + \cos \theta \bra{1} \right) \nonumber \\
\hat{\pi}_1 &=& a_1 \left(\sin \theta \ket{0} - \cos \theta \ket{1} \right) \left(\sin \theta \bra{0} - \cos \theta \bra{1} \right),
\label{genunamb}
\end{eqnarray}
chosen such that $\bra{\psi_0} \hat{\pi}_1 \ket{\psi_0} = \bra{\psi_1} \hat{\pi}_0 \ket{\psi_1} = 0$, and where $0 \leq a_0, a_1 \leq 1$.  Thus when outcome $0$ is realised, we can say for sure that the corresponding state was $\ket{\psi_0}$, while when outcome $1$ occurs, we know the state was $\ket{\psi_1}$ with certainty.  Note that these cannot form a complete measurement for any choice of $a_0$, $a_1$, unless $\ket{\psi_0}$, $\ket{\psi_1}$ are orthogonal, and thus an inconclusive outcome is needed, associated with the probability operator
\begin{equation}
\hat{\pi}_? = \hat{\mathbbmss{1}} - \hat{\pi}_0 - \hat{\pi}_1.
\label{inconc}
\end{equation}
The probability of occurrence of the inconclusive result is given by
\begin{equation}
P(?) = p_0 \bra{\psi_0} \hat{\Pi}_? \ket{\psi_0} + p_1 \bra{\psi_1} \hat{\Pi}_? \ket{\psi_1} = 1 - \sin^2 2 \theta (p_0 a_0 + p_1 a_1),
\label{probinc}
\end{equation}
and the unambiguous discrimination strategy may be further optimised by minimising this probability, subject to the constraints $a_0, a_1 \geq 0$, $\hat{\pi}_? \geq 0$.  For equal \emph{a priori} probabilities, $p_0 = p_1 = \frac{1}{2}$, the minimum value or IDP limit \cite{Ivanovic87,Dieks88,Peres88} is given by $P(?) = \cos 2 \theta = |\braket{\psi_0}{\psi_1}|$ and is achieved by the measurement
\begin{eqnarray}
\hat{\pi}_0 &=& \frac{1}{2 \cos^2 \theta} \left(\sin \theta \ket{0} + \cos \theta \ket{1} \right) \left(\sin \theta \bra{0} + \cos \theta \bra{1} \right) \nonumber \\
\hat{\pi}_1 &=& \frac{1}{2 \cos^2 \theta} \left(\sin \theta \ket{0} - \cos \theta \ket{1} \right) \left(\sin \theta \bra{0} - \cos \theta \bra{1} \right), \nonumber \\
\hat{\pi}_? &=& (1-\tan^2 \theta) \ket{0} \bra{0}.
\end{eqnarray}
For unequal prior probabilities \cite{Jaeger95}, as $p_0$ is increased, the optimal measurement is given by equations (\ref{genunamb},\ref{inconc}) with
\begin{eqnarray}
a_0 &=& \frac{1 - \sqrt{\frac{p_1}{p_0}} \cos 2 \theta}{\sin^2 2 \theta} \nonumber \\
a_1 &=& \frac{1 - \sqrt{\frac{p_0}{p_1}} \cos 2 \theta}{\sin^2 2 \theta},
\end{eqnarray}
giving $P(?) = 2 \sqrt{p_0 p_1} \cos 2 \theta$.  Thus the measurement becomes biased towards unambiguously identifying the state which is \emph{a priori} more probable.  Clearly when $\sqrt{\frac{p_0}{p_1}} \cos 2 \theta > 1$ this no longer defines a physical measurement; the optimal measurement then is simply the von Neumann measurement given by equation (\ref{vNunamb}).  In this case $\ket{\psi_1}$ always gives the inconclusive result, and the probability of failure is $P(?) = p_0 |\braket{\psi_0}{\psi_1}|^2 + p_1$.  Thus for $p_0$ much bigger than $p_1$, the optimal strategy is the one which rules out the less probable state, in contrast to the minimum error measurement, which in this regime (approximately) identifies or rules out the \emph{more} probable state.

A simple example from quantum optics might help to illustrate the main idea \cite{Hutner95}.  Let us suppose that we have an optical pulse known to have been prepared, with equal probability, in one 
the two coherent states \cite{Loudon} $\ket{\alpha}$ or $\ket{-\alpha}$.  If we interfere the pulse with a
second pulse prepared in the state $\ket{i\alpha}$ using a 50:50 symmetric beam splitter then one of
the output modes will be left in its vacuum state $\ket{0}$:
\begin{eqnarray}
\ket{\alpha}\ket{i\alpha} &\rightarrow & \ket{0}\ket{i\sqrt{2}\alpha} \nonumber \\
\ket{-\alpha}\ket{i\alpha} &\rightarrow & \ket{-\sqrt{2}\alpha}\ket{0} .
\end{eqnarray}
The state can be identified simply by detecting the light in the associated output mode.  The ambiguous outcome
is a consequence of the fact that the coherent states have a non-zero overlap with the vacuum state, and the probability for this result is
\begin{equation}
P_? = |\langle i\sqrt{2}\alpha|0\rangle|^2 = |\langle -\sqrt{2}\alpha|0\rangle|^2 = |\langle\alpha|-\alpha\rangle|,
\end{equation}
which is the IDP limit.

\subsubsection{$N > 2$ Pure States}
In the general case of discriminating unambiguously between $N$ pure states $\{ \ket{\psi_i} \}$, $i = 0, \ldots, N-1$, we wish to find probability operators $\{ \hat{\pi}_i \}$ such that
\begin{equation}
\bra{\psi_i} \hat{\pi}_j \ket{\psi_i} = P_i \delta_{ij}
\label{unambcond}
\end{equation}
where $0 \leq P_i \leq 1$.  Thus outcome $j$ is obtained only if the state is $\ket{\psi_j}$, in which case it occurs with probability $P_j$.  We first note that this is only possible if the states $\{ \ket{\psi_i} \}$ are linearly independent, as was shown by Chefles \cite{Chefles98}.  When this is the case, we can construct states $\ket{\psi_j^{\perp}}$ such that
\begin{equation}
\braket{\psi_i}{\psi_j^{\perp}} = \braket{\psi_j}{\psi_j^{\perp}} \delta_{ij},
\end{equation}
i.e. $\ket{\psi_j^{\perp}}$ is orthogonal to all allowed states except $\ket{\psi_j}$.  The POM elements
\begin{equation}
\hat{\pi}_j = \frac{P_j}{|\braket{\psi_j}{\psi_j^{\perp}}|^2} \ket{\psi_j^{\perp}} \bra{\psi_j^{\perp}}
\end{equation}
thus satisfy equation (\ref{unambcond}), and unambiguously discriminate between the linearly independent states $\{ \ket{\psi_i} \}$.  As before, an inconclusive outcome is necessary to form a complete measurement
\begin{equation}
\hat{\pi}_? = \hat{\mathbbmss{1}} - \sum_j \hat{\pi}_j.
\end{equation}
The above defines the unambiguous discrimination strategy for $N$ linearly independent states.  The occurrence of outcome $j$ indicates unambiguously that the state was $\ket{\psi_j}$.  As in the two state case, a further condition which may be applied is to minimise the probability of obtaining an inconclusive result.  Analytical solutions for the minimum achievable $P(?)$ are not known in the general case, but the solution for three states is given by Peres and Terno \cite{Peres98}, who also discuss how the method used can be extended to higher dimensions.  For the special case in which the probability of unambiguously identifying a state $\ket{\psi_j}$ is the same for all $j$ ($P_j = P, \; \forall \; j$) the minimum probability of obtaining an inconclusive result is known \cite{Chefles98}.  Further, the optimal strategy minimising this probability is given for $N$ linearly independent symmetric states in \cite{Chefles98b}.  For the general case, upper \cite{Zhang01} and lower bounds \cite{Duan98,Sun02} have been given for the probability of successful unambiguous discrimination of $N$ linearly independent states, and numerical optimisation techniques have also been considered \cite{Sun02,Eldar03b}.

\subsubsection{Mixed States}
It is only relatively recently that unambiguous discrimination has been extended to mixed states \cite{Rudolph03}, where it may be applied to problems such as quantum state comparison \cite{Barnett03,Rudolph03}, subset discrimination \cite{Zhang02}, and determining whether a given state is pure or mixed \cite{Zhang07}.  Consider the problem of discriminating between two mixed states $\hat{\rho}_0$, $\hat{\rho}_1$, which may be written in terms of their eigenvalues and eigenvectors as follows
\begin{equation}
\begin{array}{ccl}
\hat{\rho}_0 &=& \sum_i \lambda_i^{(0)} \ket{\lambda_i^{(0)}} \bra{\lambda_i^{(0)}}, \\
\hat{\rho}_1 &=& \sum_i \lambda_i^{(1)} \ket{\lambda_i^{(1)}} \bra{\lambda_i^{(1)}}.
\end{array}
\end{equation}
where $0 < \lambda_i^{(j)} \leq 1$.  Define the projectors
\begin{equation}
\begin{array}{ccl}
\hat{\Lambda}_{ker}^{(0)} &=& \hat{\mathbbmss{1}} - \sum_i \ket{\lambda_i^{(0)}} \bra{\lambda_i^{(0)}}, \\
\hat{\Lambda}_{ker}^{(1)} &=& \hat{\mathbbmss{1}} - \sum_i \ket{\lambda_i^{(1)}} \bra{\lambda_i^{(1)}},
\end{array}
\end{equation}
such that $\hat{\Lambda}_{ker}^{(0)} \hat{\rho}_0 = \hat{\Lambda}_{ker}^{(1)} \hat{\rho}_1 = 0$.  These are the projectors onto the kernels of $\hat{\rho}_0$ and $\hat{\rho}_1$ respectively\footnote{The support of a mixed state $\hat{\rho}$ is the subspace spanned by its eigenvectors with non-zero eigenvalues.  The kernel of a mixed state is the subspace orthogonal to its support.}.  If we now define $\hat{\pi}_1$ to lie in the kernel of $\hat{\rho}_0$ then $\hat{\pi}_1 = \hat{\Lambda}_{ker}^{(0)} \hat{\pi}_1 \hat{\Lambda}_{ker}^{(0)}$ and clearly
\begin{equation}
{\rm Tr}(\hat{\rho}_0 \hat{\pi}_1) = {\rm Tr}(\hat{\rho}_0 \hat{\Lambda}_{ker}^{(0)} \hat{\pi}_1 \hat{\Lambda}_{ker}^{(0)}) = 0.
\end{equation}
Thus if there exists a positive operator $\hat{\pi}_1$ in the kernel of $\hat{\rho}_0$ for which ${\rm Tr}(\hat{\rho}_1 \hat{\pi}_1) \neq 0$, then $\hat{\rho}_1$ may be unambiguously discriminated from $\hat{\rho}_0$.  Similarly $\hat{\pi}_0$ should lie in the kernel of $\hat{\rho}_1$.  Thus a necessary and sufficient condition for unambiguous discrimination between two mixed states is that they have non-identical kernels, and thus non-identical supports \cite{Rudolph03}.  Unless the states are orthogonal an inconclusive outcome will be needed, as before, $\hat{\pi}_? = \hat{\mathbbmss{1}} - \hat{\pi}_0 - \hat{\pi}_1$.  The problem of finding the strategy which minimises the probability of occurrence of the inconclusive result is again a difficult one, and one which has received much attention in the past few years.  The solutions for some special cases are known, some examples are when both states have one-dimensional kernels \cite{Rudolph03}, unambiguous discrimination between a pure and a mixed state, firstly in two dimensions \cite{Sun02b}, and later extended to $N$ dimensions \cite{Bergou03}; other examples may be found in \cite{Herzog05,Raynal07,Herzog07}.  Reduction theorems given in \cite{Raynal03} show that it is always possible to reduce the general problem to one of discriminating two states each of rank $r$, which together span a $2r$-dimensional space.  Thus the simplest case which is not reducible to pure state discrimination is the problem of two rank-2 density operators in a 4-dimensional space, which was recently analysed in detail by Kleinmann et al \cite{Kleinmann08}.  Upper and lower bounds for the general case are given in \cite{Rudolph03,Feng04,Raynal05}, a further reduction theorem in \cite{Raynal07}, and numerical algorithms are discussed in \cite{Eldar04b}.

\subsection{Maximum confidence measurements}
As pointed out in the previous section, unambiguous discrimination is possible only when the allowed states are all linearly independent.  If this is not the case, there will always be errors associated with identifying some states, even if an inconclusive outcome is allowed.  Nevertheless, we can construct an analogous measurement, one which allows us to be as confident as possible that when the outcome of measurement leads us to identify a given state $\ket{\psi_i}$, that was indeed the state prepared \cite{Croke06}.  Just as with unambiguous discrimination, this measurement is concerned with optimising the information given about the state by particular measurement outcomes, specifically the posterior probabilities
\begin{equation}
P(\hat{\rho}_i|i) = \frac{P(\hat{\rho}_i) P(i|\hat{\rho}_i)}{P(i)}.
\end{equation}
Physically, in many runs of an experiment, this probability refers to the proportion of occurrences of outcome $i$ which were due to state $\hat{\rho}_i$.  In a single-shot measurement, this therefore corresponds to the probability that it was state $\hat{\rho}_i$ that gave rise to outcome $i$.  Thus, we can think of this quantity as our confidence in taking outcome $i$ to indicate state $\ket{\psi_i}$.  In terms of the probability operator $\hat{\pi}_i$ associated with outcome $i$, we can write
\begin{equation}
P(\hat{\rho}_i|i) = \frac{p_i {\rm Tr}(\hat{\rho}_i \hat{\pi}_i)}{{\rm Tr}(\hat{\rho} \hat{\pi}_i)},
\label{conf}
\end{equation}
where $\hat{\rho} = \sum_j p_j \hat{\rho}_j$ is the \emph{a priori} density operator.  We note that $\hat{\pi}_i$ appears in both the numerator and denominator of this expression, and thus can only be determined up to a multiplicative constant.  It is always possible, therefore, to choose the overall normalisation such that
\begin{equation}
\sum_i \hat{\pi}_i \leq \hat{\mathbbmss{1}},
\end{equation}
and a physically realisable measurement may be constructed by adding an inconclusive result.  Thus the only constraint we need worry about is that $\hat{\pi}_i \geq 0$.  Optimisation of this figure of merit is greatly facilitated by the use of the ansatz
\begin{equation}
\hat{\pi}_i = \hat{\rho}^{-1/2} \hat{Q}_i \hat{\rho}^{-1/2},
\end{equation}
where, by construction, $\hat{\pi}_i \geq 0$ if $\hat{Q}_i \geq 0$.  With this, equation (\ref{conf}) becomes
\begin{equation}
P(\hat{\rho}_i|i) = {\rm Tr} \left( \hat{\rho}^{-1/2} p_i \hat{\rho}_i \hat{\rho}^{-1/2} \frac{\hat{Q}_i}{{\rm Tr}(\hat{Q}_i)} \right),
\end{equation}
where we have used the cyclical property of the trace.  Note that $\hat{Q}_i/{\rm Tr}(\hat{Q}_i)$ is a positive, trace one operator, and so has the mathematical properties of a density operator.  The density operator which has largest overlap with any operator $\hat{A}$ is simply a projector onto the largest eigenvector of $\hat{A}$ (or any density operator in the eigensubspace corresponding to the largest eigenvalue if this is degenerate).  For pure states the optimal probability operators are therefore given by
\begin{equation}
\hat{\pi}_i \propto \hat{\rho}^{-1} \hat{\rho}_i \hat{\rho}^{-1}
\label{mcpure}
\end{equation}
while for mixed states they may be written
\begin{equation}
\hat{\pi}_i \propto \hat{\rho}^{-1/2} \hat{\sigma}_i \hat{\rho}^{-1/2}
\label{mcgen}
\end{equation}
where $\hat{\sigma}_i$ is any density operator lying in the eigensubspace of $\hat{\rho}^{-1/2} p_i \hat{\rho}_i \hat{\rho}^{-1/2}$ corresponding to its largest eigenvalue.  Finally, the limit is given by
\begin{equation}
[P(\hat{\rho}_i|i)]_{max} = \gamma_{max} \left( \hat{\rho}^{-1/2} p_i \hat{\rho}_i \hat{\rho}^{-1/2} \right)
\label{maxconf}
\end{equation}
where $\gamma_{max}(\hat{A})$ denotes the largest eigenvalue of $\hat{A}$.

\begin{figure}[h]
\begin{center}
\includegraphics[width=120truemm]{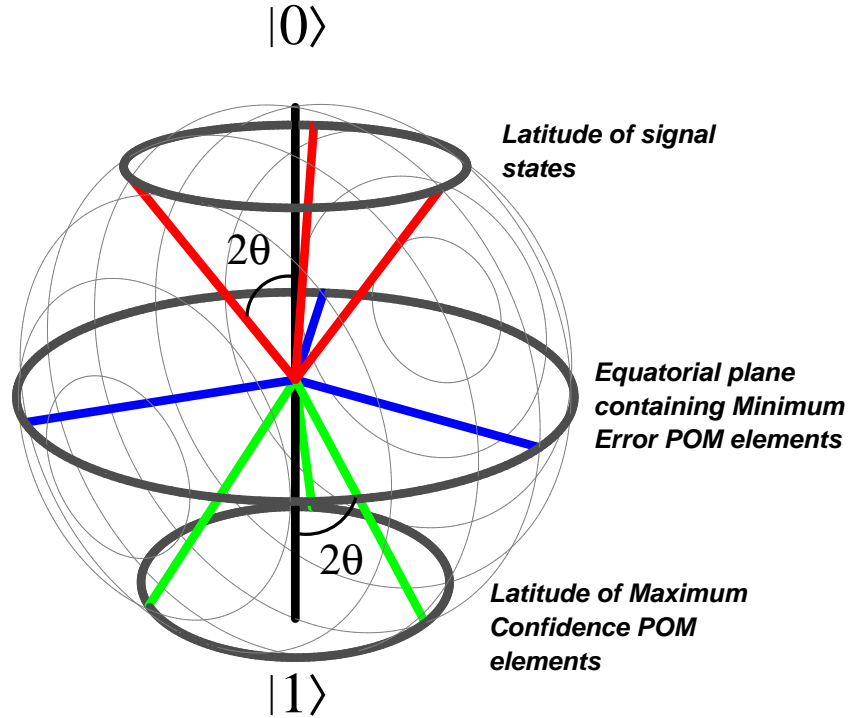}
\end{center}
\caption{Bloch sphere representation of states.  The states used in the example, along with the states onto which the optimal maximum confidence and minimum error POM elements project are shown. (\cite{Croke06}, Copyright (2006) by the American Physical Society.)}
\label{bloch}
\end{figure}

The simplest non-trivial example of a set of linearly dependent states is that of three states in two dimensions.  To illustrate this strategy we consider the problem of discriminating between the states
\begin{eqnarray}
\label{states}
\ket{\psi_0} & = &\cos\theta\ket{0} + \sin\theta\ket{1}, \nonumber \\
\ket{\psi_1} & = & \cos\theta\ket{0} + e^{2 \pi i / 3} \sin\theta\ket{1}, \\
\ket{\psi_2} & = & \cos\theta\ket{0} + e^{-2 \pi i / 3} \sin\theta\ket{1}, \nonumber
\end{eqnarray}
where $0 \leq \theta \leq \pi/4$, occurring with equal $\emph{a priori}$ probabilities $p_i = 1/3, i=0,1,2$.  These states are symmetrically located at the same latitude of the Bloch sphere, as shown in Fig. \ref{bloch}.  The \emph{a priori} density operator for this set is
\begin{equation}
\hat{\rho} = \cos^2 \theta \ket{0} \bra{0} + \sin^2 \theta \ket{1} \bra{1},
\end{equation}
and the maximum confidence POM elements may be readily calculated using equation (\ref{mcpure}).  These have the form $\hat{\pi}_i = \alpha_i \ket{\phi_i} \bra{\phi_i}$, where we have some freedom in choosing the constants $\alpha_i, i=0,1,2$, and
\begin{eqnarray}
\label{MCPOM}
\ket{\phi_0} & =  &\sin\theta\ket{0} + \cos\theta\ket{1}, \nonumber \\
\ket{\phi_1} & = & \sin\theta\ket{0} + e^{2 \pi i / 3} \cos\theta\ket{1}, \\
\ket{\phi_2} & = & \sin\theta\ket{0} + e^{-2 \pi i / 3} \cos\theta\ket{1}. \nonumber
\end{eqnarray}
These states correspond to reflections of the input states in the equatorial plane of the Bloch sphere, and are also shown in Fig. \ref{bloch}.  It is not possible in general to choose $\alpha_0, \alpha_1, \alpha_2$ such that $\{ \hat{\pi}_i \}$ form a complete measurement, and thus an additional operator, $\hat{\pi}_? = \hat{\mathbbmss{1}} - \sum_i \hat{\pi}_i$, associated with an inconclusive result, is needed.  We may choose to complete the measurement by minimising the probability of an inconclusive result
\begin{equation}
\begin{array}{ccl}
P(?) &=& {\rm Tr}(\hat{\rho} \hat{\pi}_?) \\
&=& 1 - 2(\alpha_0 + \alpha_1 + \alpha_2) \cos^2 \theta \sin^2 \theta.
\end{array}
\end{equation}
As $P(?)$ is a monotonically decreasing function of $\alpha_i$, the optimal values of these parameters lie on the boundary of the allowed domain, defined by the constraint $\hat{\pi}_? \geq 0$.  It is straightforward to show that this leads us to choose $\alpha_0 = \alpha_1 = \alpha_2 = (3 \cos^2 \theta)^{-1}$, giving
\begin{equation}
\label{Piinc}
\hat{\pi}_? = (1-\tan^2 \theta) \ket{0} \bra{0}.
\end{equation}

\begin{figure}[h]
\begin{center}
\includegraphics[width=120truemm]{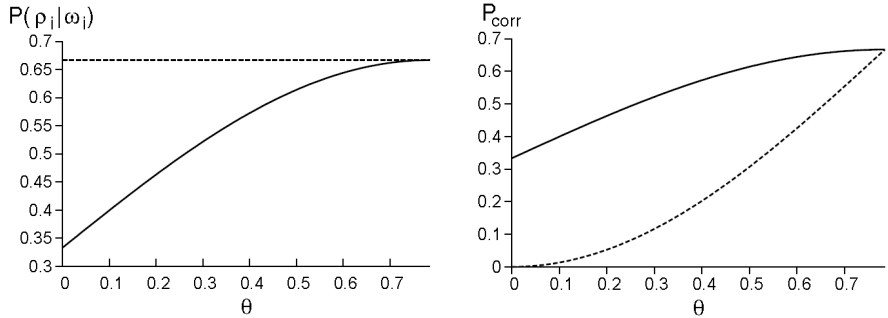}
\end{center}
\caption{Graphs showing the maximum confidence (left) and minimum error (right) figures of merit, for various values of the parameter $\theta$ for the example discussed in the text.  In each case the values achieved by the optimal maximum confidence measurement are indicated by a dashed line and those corresponding to the optimal minimum error measurement are indicated by a solid line.}
\label{fom}
\end{figure}

It is useful to compare this measurement with the minimum error (ME) measurement, which for this set is given by the square root measurement discussed earlier 
\begin{equation}
\label{MEPOM}
\hat{\pi}_i^{ME} = \frac{1}{3} \hat{\rho}^{-1/2} \ket{\psi_i} \bra{\psi_i} \hat{\rho}^{-1/2} = \frac{2}{3} \ket{\phi_i^{ME}} \bra{\phi_i^{ME}},
\end{equation} where
\begin{eqnarray}
\label{MEstates}
\ket{\phi_0^{ME}} & =  &\frac{1}{\sqrt{2}}(\ket{0} + \ket{1}), \nonumber \\
\ket{\phi_1^{ME}} & = & \frac{1}{\sqrt{2}}(\ket{0} + e^{2 \pi i / 3} \ket{1}), \\
\ket{\phi_2^{ME}} & = & \frac{1}{\sqrt{2}}(\ket{0} + e^{-2 \pi i / 3} \ket{1}). \nonumber
\end{eqnarray}
In the Bloch sphere representation, these states correspond to the projection of the input states onto the equatorial plane, as can be seen in Fig. \ref{bloch}.  The minimum error and maximum confidence figures of merit are shown for each measurement in Fig. \ref{fom}.  For the minimum error measurement, each outcome leads us to identify some state, and the average probability of making an error is minimised.  However, the confidence in identifying a state may be increased by allowing for an inconclusive result, as may be seen from the plots.  When a non-inconclusive result is obtained in the maximum confidence measurement, the probability that the state prepared really was the one identified is $\frac{2}{3}$, compared with $\frac{1}{3}(1+\sin 2 \theta)$ for the minimum error measurement.

\subsubsection{Other Similar Strategies}
A related strategy may be constructed by applying a worst case optimality criterion to the conditional probability considered here, $P(\hat{\rho}_i|i)$ \cite{Kosut04}.  This approach does not allow for inconclusive results, but searches for the measurement for which the smallest value of $P(\hat{\rho}_i|i)$ is maximised.  This more complicated problem is difficult to solve analytically, but may be cast as a quasi-convex optimisation, for which efficient numerical techniques are available.  An alternative strategy allows inconclusive results to occur with a certain fixed probability, $P_I$, and maximises the probability of correctly identifying the state when a non-inconclusive outcome is obtained.  For linearly independent pure states this approach interpolates between minimum error and unambiguous discrimination \cite{Chefles98c,Zhang99}.  For mixed states a similar approach is possible \cite{Fiurasek03}, which may be interpreted as interpolating between a minimum error measurement and a maximum confidence strategy.  It is clear that the probability of obtaining a correct result, renormalised over only the results which are not inconclusive, denoted $P_{RC}$, can never be larger than the largest value of $P(\hat{\rho}_i|i)_{max}$ for a given set, regardless of how much we increase $P_I$.  This upper bound is achieved by a maximum confidence strategy which only ever identifies the state(s) $\hat{\rho}_i$ such that (from equation (\ref{maxconf}))
\begin{equation}
\gamma_{max} \left( \hat{\rho}^{-1/2} p_i \hat{\rho}_i \hat{\rho}^{-1/2} \right) \geq \gamma_{max} \left( \hat{\rho}^{-1/2} p_j \hat{\rho}_j \hat{\rho}^{-1/2} \right) \quad \forall \; \hat{\rho}_j,
\end{equation}
while all other results are interpreted as inconclusive.  Although it is difficult to find the optimal measurement for general $P_I$, it is indeed found that $P _{RC}$ is saturated at some value of $P_I$, and the maximum $P_{RC}$ achievable corresponds to the strategy outlined here \cite{Fiurasek03}.

\subsubsection{Related problems - quantum state filtration}
Quantum state filtration refers to the problem of whether the state of a system is a given state $\ket{\psi_i}$ or simply in any one of the other states in a given set $\{ \ket{\psi_j} \}$, $j \neq i$.  This problem is less demanding than complete discrimination among all possible states, and in the minimum error approach the probability of error may be smaller in the state filtration case \cite{Herzog02}.  For the maximum confidence measurement however, the optimality of the probability operator $\hat{\pi}_i$ in equation (\ref{mcgen}) is independent of the number and interpretation of other possible outcomes.  Thus the confidence in identifying a given state from a set cannot be increased by considering this simpler problem.  This figure of merit is dependent only on the geometry of the set, and in this sense can be thought of as a measure of how distinguishable $\hat{\rho}_i$ is in the given set.

\subsection{Comments on the Relationships Between Strategies}
The maximum confidence strategy was introduced as an analogy to unambiguous discrimination for linearly dependent states \cite{Croke06}.  In fact, unambiguous discrimination is a special case of maximum confidence discrimination.  The maximum confidence measurement maximises the conditional probability $P(\hat{\rho}_i|i)$.  If this figure of merit is equal to unity for some state $\hat{\rho}_i$, the optimal measurement is such that, when outcome $i$ is obtained, we can be absolutely certain that $\hat{\rho}_i$ was in fact the state received, corresponding to unambiguous discrimination.  We can use the maximum confidence formalism to investigate when unambiguous discrimination is possible.  Equation (\ref{conf}) may be written
\begin{equation}
P(\hat{\rho}_i|i) = \frac{p_i {\rm Tr}(\hat{\rho}_i \hat{\pi}_i)}{p_i {\rm Tr}(\hat{\rho}_i \hat{\pi}_i) + \sum_{j \neq i} p_j {\rm Tr}(\hat{\rho}_j \hat{\pi}_i)}
\end{equation}
Clearly the limit of unity may be achieved if there exists any projector $\hat{\Lambda}_i$ for which $\hat{\Lambda}_i \sum_{j \neq i} p_j \hat{\rho}_j \hat{\Lambda}_i = 0$ while $\hat{\Lambda}_i \hat{\rho}_i \hat{\Lambda}_i$ is non-zero.  $\hat{\pi}_i$ is then any operator lying in the subspace with projector $\hat{\Lambda}_i$.  This reproduces the known results that unambiguous discrimination is possible between pure states if the states are linearly independent \cite{Chefles98}, and between mixed states if they have distinct supports \cite{Rudolph03}.  More precisely, a measurement is possible which will sometimes allow us to identify $\hat{\rho}_i$ unambiguously if $\hat{\rho}_i$ has support in the kernel of $\sum_{j \neq i} p_j \hat{\rho}_j$.  This condition is less restrictive than the previous, which does not hold in the case where it is possible to unambiguously discriminate some but not all states in a set.  Unambiguous discrimination is still possible in this case, but some states are never identified.  For example, it was pointed out by Sun et al \cite{Sun02b}, that it is possible to apply unambiguous discrimination to the problem of determining whether a system is in a given state $\ket{\psi_0}$ or in either of two other possible states, $\ket{\psi_1}$, $\ket{\psi_2}$, even if the states span only two dimensions, and therefore are linearly dependent.  This may be more easily understood as unambiguous discrimination between a mixed state and a pure state in two dimensions \cite{Rudolph03}.  Let
\begin{eqnarray}
\rho_{0} &=&  \ket{\psi_0} \bra{\psi_0}, \nonumber \\
\rho_1 &=& \frac{p_1}{p_1+p_2} \ket{\psi_1} \bra{\psi_1} + \frac{p_2}{p_1+p_2} \ket{\psi_2} \bra{\psi_2} = q \ket{0} \bra{0} + (1-q) \ket{1} \bra{1},
\end{eqnarray}
where $\ket{0}$, $\ket{1}$ are the eigenkets of $\rho_1$, $0 < q < 1$, and without loss of generality we can write $\ket{\psi_0} = \cos \theta \ket{0} + \sin \theta \ket{1}$.  It is clear that the von Neumann measurement
\begin{eqnarray}
\hat{\pi}_0 &=& \ket{\psi_0} \bra{\psi_0} \nonumber \\
\hat{\pi}_1 &=& \hat{\mathbbmss{1}} - \ket{\psi_0} \bra{\psi_0}
\end{eqnarray}
can unambiguously discriminate the two possibilities -if outcome $1$ is obtained, we can say for sure that the state was $\rho_1$, while the result $0$ is interpreted as inconclusive.  However this measurement never tells us if the state was $\ket{\psi_0}$.  In this case it may be useful to consider unambiguous discrimination within the framework of maximum confidence measurements.  It is then possible to construct a measurement which sometimes identifies $\hat{\rho}_1$ with certainty, but also sometimes identifies $\hat{\rho}_0$ as confidently as possible.  In general an inconclusive result will also be necessary.

Now suppose that instead of maximising the conditional probability in equation (\ref{conf}) independently for each state in the set we choose to maximise a weighted average of these probabilities.  We would then obtain as our figure of merit
\begin{equation}
P(\hat{\rho}_i|i)_{avg} = \sum_i P(i) P(\hat{\rho}_i| i) = \sum_i P(\hat{\rho}_i) P(i|\hat{\rho}_i),
\label{mefom}
\end{equation}
which is precisely the figure of merit maximised by the minimum error measurement.  Thus these two strategies can be thought of as applying a different optimality condition to the same quantity.  The minimum error measurement also has the additional constraint that the operators $\{\hat{\pi}_i \}$ must form a complete measurement, as it is never optimal to allow inconclusive results to occur.  This constraint makes finding the optimal measurement a difficult problem, although the conditions which the optimal measurement must satisfy are known, as we have shown.  By contrast, the maximum confidence strategy allows a closed form solution for an arbitrary set of states.  In the special case where the maximum confidence figure of merit is the same for all states $\hat{\rho}_i$ and no inconclusive result is needed, the two strategies coincide.  More generally, it is clear by examination of equation (\ref{mefom}) that an upper bound for the minimum error figure of merit is given by the largest value of $P(\hat{\rho}_i|i)_{max}$ for a given set (i.e. the largest value of equation (\ref{maxconf})).

\subsection{Mutual information}

In communications theory the performance of a communications channel is quantified not by 
an error probability but rather by the information conveyed.  We can give a precise meaning
to this by invoking Shannon's noisy channel coding theorem \cite{Shannon,Cover}, which states 
that the maximum communications rate, or channel capacity, is obtained by maximising the 
mutual information between the transmitter and receiver.  If the transmitted message, $A$, is
one of the set $\{a_i\}$ and the reception event, $B$, is one of the set $\{b_j\}$, then the 
mutual information is defined to be
\begin{equation}
\label{InfDef}
H(A:B) = \sum_{ij}P(a_i,b_j)\log\left(\frac{P(a_i,b_j)}{P(a_i)P(b_j)}\right) ,
\end{equation}
where the logarithm is usually taken to be base 2 so that the information is expressed in bits.
For a quantum channel, the state $\hat\rho_i$ is selected with probability $p_i$ and the measurement
result $b_j$ is associated with the probability operator $\hat\pi_j$.  It follows that the mutual 
information is 
\begin{equation}
\label{QMutInf}
H(A:B) = \sum_{ij} p_i{\rm Tr}\left(\hat\rho_i\hat\pi_j\right)\log\left(
\frac{{\rm Tr}\left(\hat\rho_i\hat\pi_j\right)}{{\rm Tr}\left(\hat\rho\hat\pi_j\right)}\right) ,
\end{equation}
where $\hat\rho = \sum_ip_i\hat\rho_i$.  The maximum value of the mutual information is
found by varying both the preparation probabilities, $p_i$ and the measurement strategies.
This is a very difficult optimisation problem and there are very few exact solutions known
\cite{Davies,Levitin}.  A scarcely simpler problem is to fix the preparation probabilities and
then seek the maximum value to give what is referred to as the accessible information
\cite{Sasaki99}.

For two pure states, it is known that the mutual information is maximised if the states are 
prepared with equal probability and if the minimum error measurement is employed
\cite{Levitin}.  For three or more states, the accessible information is known if the states
are equally likely to be selected and possess a degree of symmetry.  In particular, for the 
so-called trine ensemble of three equally probable states (\ref{TrineDef}), the accessible
information is obtained with a generalised measurement with probability operators
\begin{eqnarray}
\hat\pi_0 &=& \frac{2}{3}|1\rangle\langle 1| \nonumber \\
\hat\pi_1 &=&  \frac{2}{3}\left(\frac{1}{2}|1\rangle + \frac{\sqrt 3}{2}|0\rangle\right)
\left(\frac{1}{2} \langle1| + \frac{\sqrt 3}{2}\langle0|\right)  \nonumber \\
\hat\pi_2 &=&  \frac{2}{3}\left(\frac{1}{2}|1\rangle - \frac{\sqrt 3}{2}|0\rangle\right)
\left(\frac{1}{2} \langle1| - \frac{\sqrt 3}{2}\langle0|\right) .
\end{eqnarray}
Note that the accessible information is obtained not by maximising the probability for
determining the state but rather for \emph{eliminating} one of the states so that
\begin{equation}
\langle \psi_i|\hat\pi_j|\psi_i\rangle = \frac{1}{2}(1 - \delta_{ij}).
\end{equation}
A similar strategy works well for four equiprobable states arranged so as to form a regular
tetrahedron on the Bloch or Poincar\'e sphere \cite{Davies}.  For more states,
optimal strategies have been demonstrated with fewer measurement outcomes than states
\cite{Sasaki99}.

\subsection{No signaling bounds on state discrimination}
Up to now we have discussed the limits on quantum state discrimination by mathematically formulating figures of merit which may then be evaluated and compared for any allowed measurement by virtue of the generalised measurement formalism.  It is interesting to note however that it is possible to place tight bounds on state discrimination without any reference to generalised measurements, by appealing to the no signaling principle, the condition that information may not propagate faster than the speed of light.

Although entanglement appears to allow space-like separated quantum systems to influence one another instantaneously, it may be shown that quantum mechanical correlations do not allow signaling \cite{Ghirardi80,Bussey82,Jordan83,Bussey87}.  Further, due to the implications of this in reconciling quantum mechanics with special relativity, it has been suggested that the no-signaling principle be given the status of a physical law, which may be used to limit quantum mechanics and possible extensions of it \cite{Popescu94,Masanes06}.  In practice, bounds on the fidelity of quantum cloning machines \cite{Gisin98,Ghosh99}, the success probability of unambiguous discrimination \cite{Barnett02,Feng02}, and the maximum confidence figure of merit \cite{Croke08} have been derived using no-signaling arguments.  In particular, the no-signaling principle may be used to put a tight bound on unambiguous discrimination of two pure states \cite{Barnett02}, and to derive the maximum confidence strategy \cite{Croke08}.  We will discuss these two cases here.

\subsubsection{Unambiguous discrimination}
Consider the entangled state
\begin{equation}
\ket{\Psi} = \sqrt{p_0} \ket{\psi_0}_L \ket{0}_R + \sqrt{1-p_0} \ket{\psi_1}_L \ket{1}_R
\end{equation}
where $\ket{\psi_0}_L$, $\ket{\psi_1}_L$ are non-orthogonal states of the left system (given by equation (\ref{Twostates})), and $\ket{0}_R$, $\ket{1}_R$ forms an orthonormal basis for the right system.  The reduced density operator of the right system may be obtained by taking the partial trace over the left system, and is given by
\begin{equation}
\hat{\rho}_R = {\rm Tr}_L \left( \ket{\Psi} \bra{\Psi} \right) = \left( \begin{array}{cc} p_0 & \sqrt{p_0(1-p_0)} \cos 2 \theta \\ \sqrt{p_0(1-p_0)} \cos 2 \theta & 1-p_0 \end{array} \right).
\end{equation}
According to the no-signaling principle, no operation performed on the left system may be detected by measurement of the right system alone, as this could be used to signal faster than light.  Thus, after any physically allowed transformation of the left system, the reduced density operator of the right system must remain the same.  Consider now a measurement which discriminates unambiguously between the states $\ket{\psi_0}_L$, $\ket{\psi_1}_L$ of the left system.  If outcome 0 is realised, which occurs with some probability $q_0$, the right system is projected into the state $\ket{0}_R$, due to the inital entanglement between the systems.  Similarly outcome 1 projects the right system into state $\ket{1}_R$, with probability $q_1$.  There is also the inconclusive result, which transforms the right system to some as yet unknown state 
\begin{equation}
\hat{\rho}_? = \left( \begin{array}{cc} \rho_?^{00} & \rho_?^{01} \\ \rho_?^{10} & \rho_?^{11} \end{array} \right)
\end{equation}
with probability $q_?$.  No signaling implies
\begin{equation}
\hat{\rho}_R = \left( \begin{array}{cc} q_0 & 0 \\ 0 & q_1 \end{array} \right) + q_? \left( \begin{array}{cc} \rho_?^{00} & \rho_?^{01} \\ \rho_?^{10} & \rho_?^{11} \end{array} \right).
\end{equation}
The task is then to minimise $q_?$ subject to the above condition and the conditions $\hat{\rho}_? \geq 0$, $q_0, q_1, q_? \geq 0$.  This optimisation is straight-forward \cite{Barnett02}, and remarkably gives precisely the Jaeger and Shimony result \cite{Jaeger95} discussed in Section 3.2.  Thus the no-signaling condition may be used to place a tight bound on the success probability of unambiguous discrimination, without any reference to generalised measurements.

\subsubsection{Maximum confidence measurements}
The confidence in identifying a given state $\ket{\psi_j}$ as a result of a state discrimination measurement on the ensemble $\{ \ket{\psi_i}, p_i \}$ is simply the probability that it was state $\ket{\psi_j}$ that gave rise to the measurement outcome observed.  Consider now the entangled state
\begin{equation}
\ket{\Psi} = \sum_{i=0}^{N-1} \sqrt{p_i} \ket{\psi_i}_L \ket{i}_R
\end{equation}
where $\{ \ket{\psi_i}_L \}$ are non-orthogonal states of the left system which together span a $D \leq N$-dimensional space, and $\{ \ket{i}_R \}$ form an orthonormal basis for the right system.  Now for any measurement performed on the left system of the entangled pair, the probability that it was state $\ket{\psi_j}_L$ which gave rise to the observed outcome is equivalent to the probability that the right system is now found in state $\ket{j}_R$.  Thus, if measurement outcome $j$ causes the right system to transform to $\hat{\rho}_{R|j}$, we can write
\begin{equation}
P(\psi_j|j) = {}_R \bra{j} \hat{\rho}_{R|j} \ket{j}_{R}.
\end{equation}
It may be shown (by reference to the Schmidt decomposition of $\ket{\Psi}$ \cite{Nielsen}), that although the right system lies in an $N$-dimensional Hilbert space, it is confined to a $D$-dimensional subspace (with projector denoted $\hat{P}_D$ below) due to the entanglement with the left system.  The key point then, is to notice that any operation performed on the left system cannot take the right system out of this subspace, since this could be detected with some probability by a measurement on the right system alone, and thus could be used to signal.  Thus ${}_R \bra{j} \hat{\rho}_{R|j} \ket{j}_{R}$ is restricted by the requirement that $\hat{\rho}_{R|j}$ lies in this subspace, and is clearly bounded by the magnitude of the projection of $\ket{j}_R$ onto this space
\begin{equation}
P(\psi_j|j) = {}_R \bra{j} \hat{\rho}_{R|j} \ket{j}_{R} \leq {}_R \bra{j} \hat{P}_D \ket{j}_{R}.
\end{equation}
Further, this bound is achievable and is equivalent to that obtained previously (equation (\ref{maxconf})) \cite{Croke08}.  Similar arguments may be applied to the mixed state case, and the maximum confidence strategy is derived in a natural way from no-signaling considerations.  Finally, we note that in the case where the states $\{ \ket{\psi_i}_L \}$ are linearly independent, $D=N$, and the right system occupies the entire $N$-dimensional Hilbert space.  In this case the limit is unity, corresponding to unambiguous discrimination.

\section{State Discrimination - Experiments}

The theory of generalised measurements has a mathematically appealing generality in that it
depends only on the overlaps of the possible states to be discriminated and on the probabilities
that each was the state prepared.  The nature of the physical states be they nuclear spins, optical 
coherent states or electronic energy levels in an atom, is unimportant.  In performing experimental
demonstrations, however, the choice of physical system is of primary importance.  We require a
physical system in which superpositions are relatively stable, easy to prepare and to manipulate and
also, of course, to measure.  For all these reasons, the system of choice has usually been photon
polarisation and forms the basis of our review.

\subsection{Photon Polarisation}

\begin{figure}[h]
\begin{center}
\includegraphics[width=120truemm]{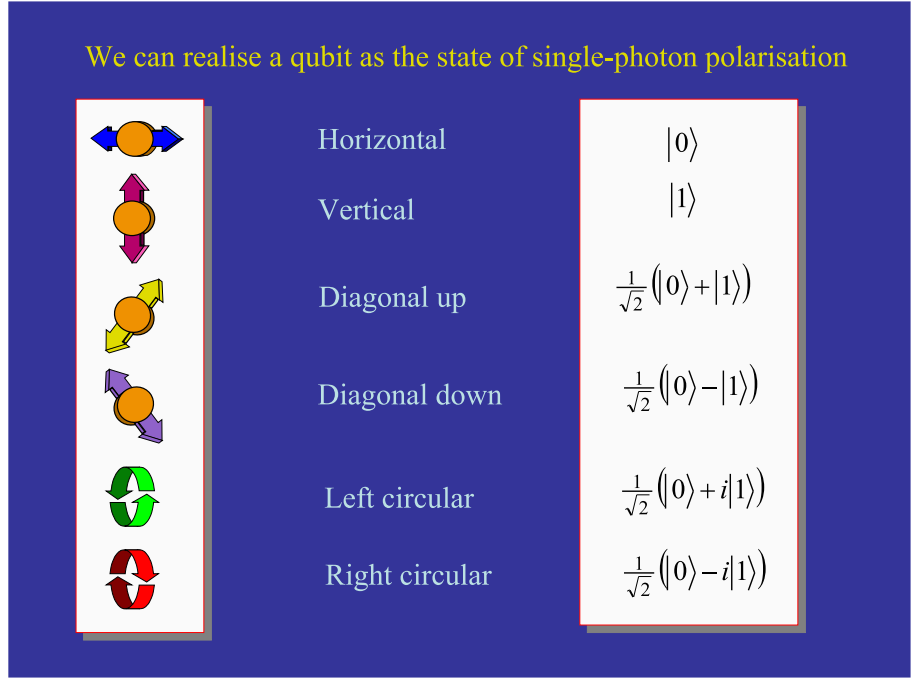}
\end{center}
\caption{Polarisation of light as a two-level system, or qubit.}
\label{polqubit}
\end{figure}

At least within paraxial optics \cite{Siegman} the electric and magnetic fields are very nearly
perpendicular to the direction of propagation of the light.  It is conventional to define the 
polarisation by the orientation of the electric field in this transverse plane \cite{Fowles}.
Two orthogonal polarisations then correspond to fields in which the electric fields are 
oriented at $90^\circ$ to each other.   The polarisation of a single photon is an excellent
two-state quantum system, or qubit \cite{Barnett09,Nielsen} as we can identify the states of
horizontal and vertical polarisation with the logical $|0\rangle$ and $|1\rangle$ states of a
qubit:
\begin{equation}
\label{Eq4.1.1}
|0\rangle = |H\rangle \qquad |1\rangle = |V\rangle .
\end{equation}
Other polarisations are superpostions of these states.  In particular, as illustrated in Fig. \ref{polqubit}, linear polarisation at $\pm 45^\circ$ to the horizontal and circular polarisations are the superpositions
\begin{eqnarray}
\label{Eq4.1.2}
|+45^\circ\rangle &=& \frac{1}{\sqrt 2}\left(|0\rangle + |1\rangle\right) \nonumber \\
|-45^\circ\rangle &=& \frac{1}{\sqrt 2}\left(|0\rangle - |1\rangle\right) \nonumber \\
|L \rangle &=& \frac{1}{\sqrt 2}\left(|0\rangle + i|1\rangle\right) \nonumber \\
|R \rangle &=& \frac{1}{\sqrt 2}\left(|0\rangle - i|1\rangle\right) .
\end{eqnarray}
The set of all possible pure states of polarisation can be represented on the surface of a sphere,
the Poincar\'e sphere \cite{Born,Wolf}, which is an equivalent representation to the Bloch sphere
used for qubits in quantum information theory \cite{Barnett09,Nielsen}.
States of optical polarisation can be changed coherently by delaying one polarisation compared with the orthogonal polarisation, usually by a quarter or half a wavelength, using birefringent wave plates.  A combination of three suitably oriented half- and quarter-wave plates can 
perform any desired transformation, corresponding to a rotation on the Poincar\'e sphere through
any desired angle about any desired axis.  In this way we can realise any desired single-qubit
unitary transformation.

\begin{figure}[h]
\begin{center}
\includegraphics[width=65truemm]{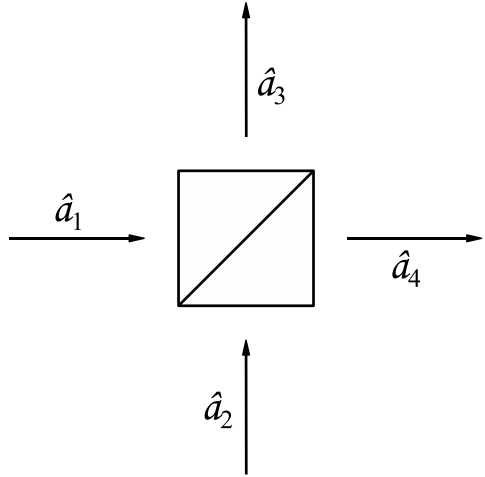}
\end{center}
\caption{A beamsplitter can be used to superpose or separate field modes.  The input and output modes are labeled with the associated annihilation operators.}
\label{beamsplitter}
\end{figure}
It is important, in order to realise generalised measurements, to be able to superpose fields
and also to be able to spatially separate different polarisations.  These tasks can be performed 
using beam-splitters and polarising beam splitters.  For fully overlapping modes with the same
frequency, we can write the output annihilation operators in terms of those for the input modes.
For a symmetric polarisation-independent beam splitter we find \cite{Loudon}
\begin{eqnarray}
\label{Eq4.1.3}
\hat a_3^{H,V} = r \hat a_1^{H,V} + t \hat a_2^{H,V} \nonumber \\
\hat a_4^{H,V} = t \hat a_1^{H,V} + r \hat a_2^{H,V},
\end{eqnarray}
where the input and output modes are labelled as in Fig. \ref{beamsplitter}.

Enforcing the canonical commutation relations at for the output modes constrains the 
reflection and transmission coefficients:
\begin{equation}
\label{Eq4.1.4}
|t|^2 + |r|^2 = 1, \qquad rt^* + tr^* = 0.
\end{equation}
A polarising beam splitter is designed to transmit horizontally polarised light and to reflect 
vertically polarised light.  This means that input and output annihilation operators are related by
\begin{eqnarray}
\label{Eq4.1.5}
\hat a_3^H = \hat a_2^H  \qquad \hat a_3^V = \hat a_1^V  \nonumber \\
\hat a_4^H = \hat a_1^H  \qquad \hat a_4^V = \hat a_2^V .
\end{eqnarray}
In correlating photon polarisation and direction, a polarising beam splitter can be used to
prepare (filter) light with a desired polarisation or, in conjunction with photodetectors placed
in each output beam, to measure the polarisation.  They also allow us to perform different transformations on two orthogonal polarisations and this is crucial in enabling us to perform
generalised measurements.

We should make one important point before describing any of the experiments that have been
performed and this is that they have not been done with single photon sources.  All of them rely
on linear optical elements and processes and for these, the single-photon probability amplitudes
and the associated probabilities behave in the same way as the amplitudes and intensities
of classical optics.  Some of the experiments have been performed at light levels in the quantum
regime, however, and this suggests strongly that the devices will work in the same way given 
single photon sources and detectors.

\subsection{Minimum Error Discrimination}

\subsubsection{Two states}
\begin{figure}[h]
\begin{center}
\includegraphics[width=120truemm]{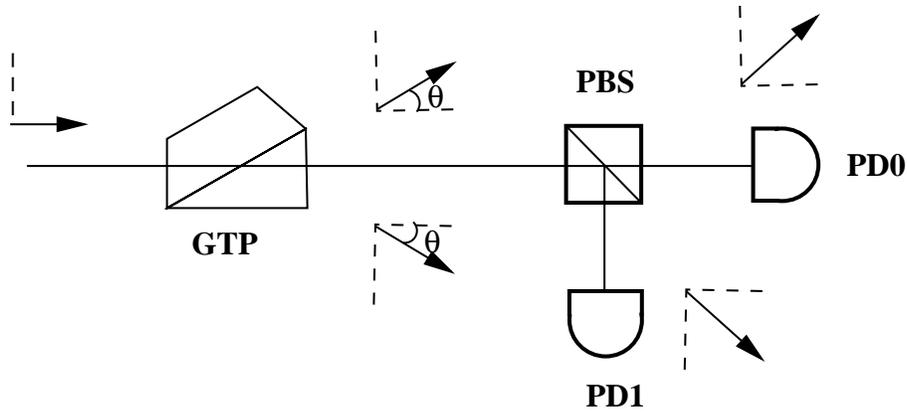}
\end{center}
\caption{Schematic of the Barnett-Riis experiment achieving the Helstrom bound for state discrimination between two pure states.}
\label{minerrexpt}
\end{figure}
The simplest minimum error problem is, as we have seen, that for two pure states (\ref{Twostates}).
 For the photon polarisations described above these correspond to two states of linear
 polarisation, oriented at $+\theta$ and $-\theta$ to the horizontal, so that the angle between them
 is $2\theta$, for a range of values of $\theta$ between $0$ and $\pi/4$.  If the two states are prepared with equal prior probability then, as we have seen, the
 minimum error measurement corresponds to a familiar von Neumann, or projective, measurement
 with two projectors associated with the orthogonal states (\ref{Twomeas}).  For optical polarisation, 
 this corresponds to measuring the polarisation at $45^\circ$ to the horizontal.  Thus the minimum error strategy in this case is a simple polarisation measurement.
 The experiment to test this \cite{Riis} was performed using light pulses with on average 0.1 photons 
 per pulse prepared in the desired polarisation state by use of a Glan-Thompson polariser oriented
 so as to produce polarised light at the angle $+\theta$ or $-\theta$ to the horizontal.  These were then
 measured using a polarising beam splitter oriented so as to transmit light polarised at $+45^\circ$ to the
 horizontal and to reflect the orthogonal polarisation.  The experimental apparatus is shown in Fig. \ref{minerrexpt}.  Results were found to be in excellent agreement with the Helstrom value (\ref{PHelstrom}) for equal prior probabilities:
 \begin{equation}
 \label{Eq4.2.1}
 P_{err} = \frac{1}{2}\left(1 - \sin 2\theta\right) .
 \end{equation}
 
 \subsubsection{Three or four states}
 
\begin{figure}[h]
\begin{center}
\includegraphics[width=60truemm]{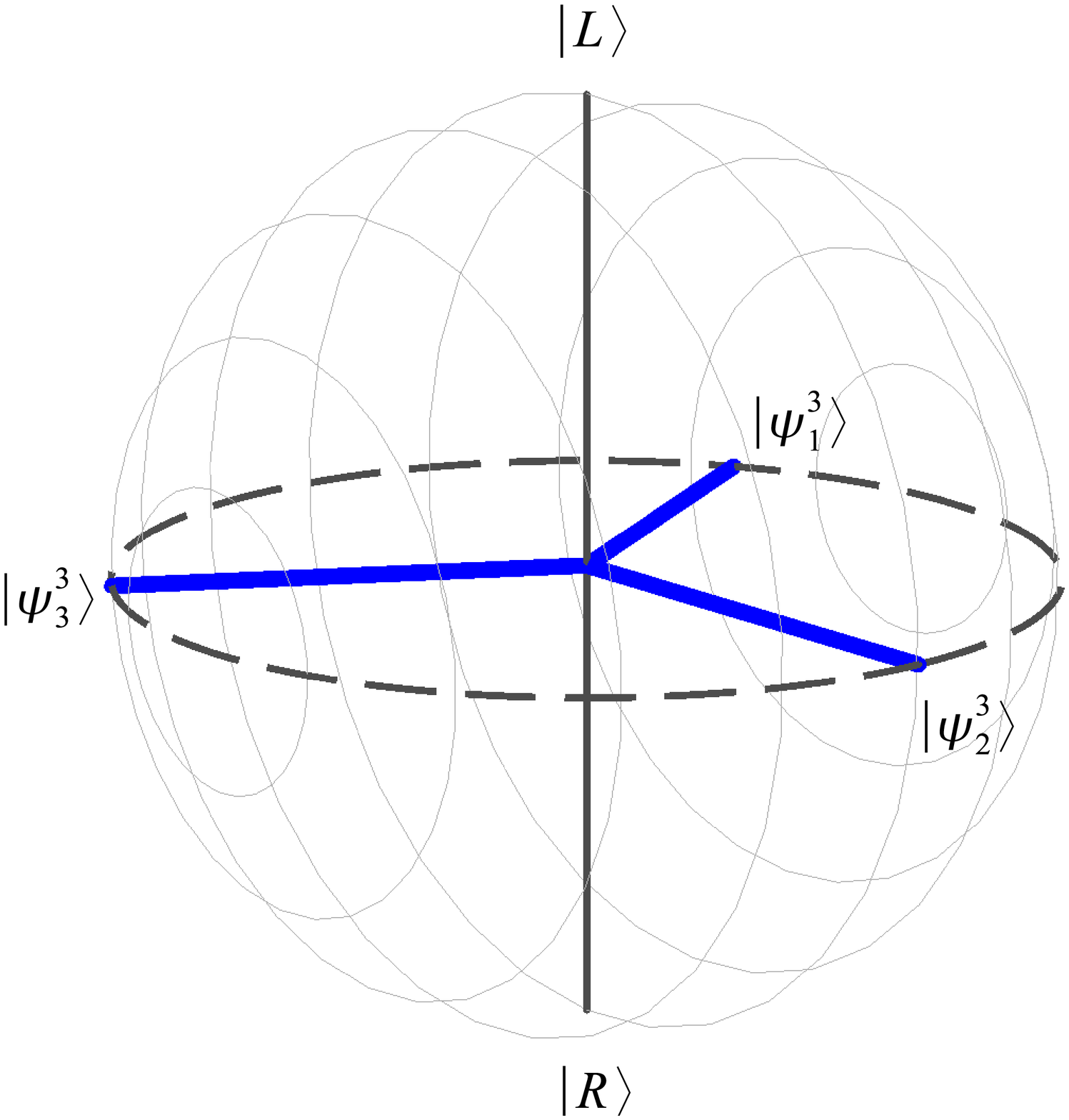}
\includegraphics[width=65truemm]{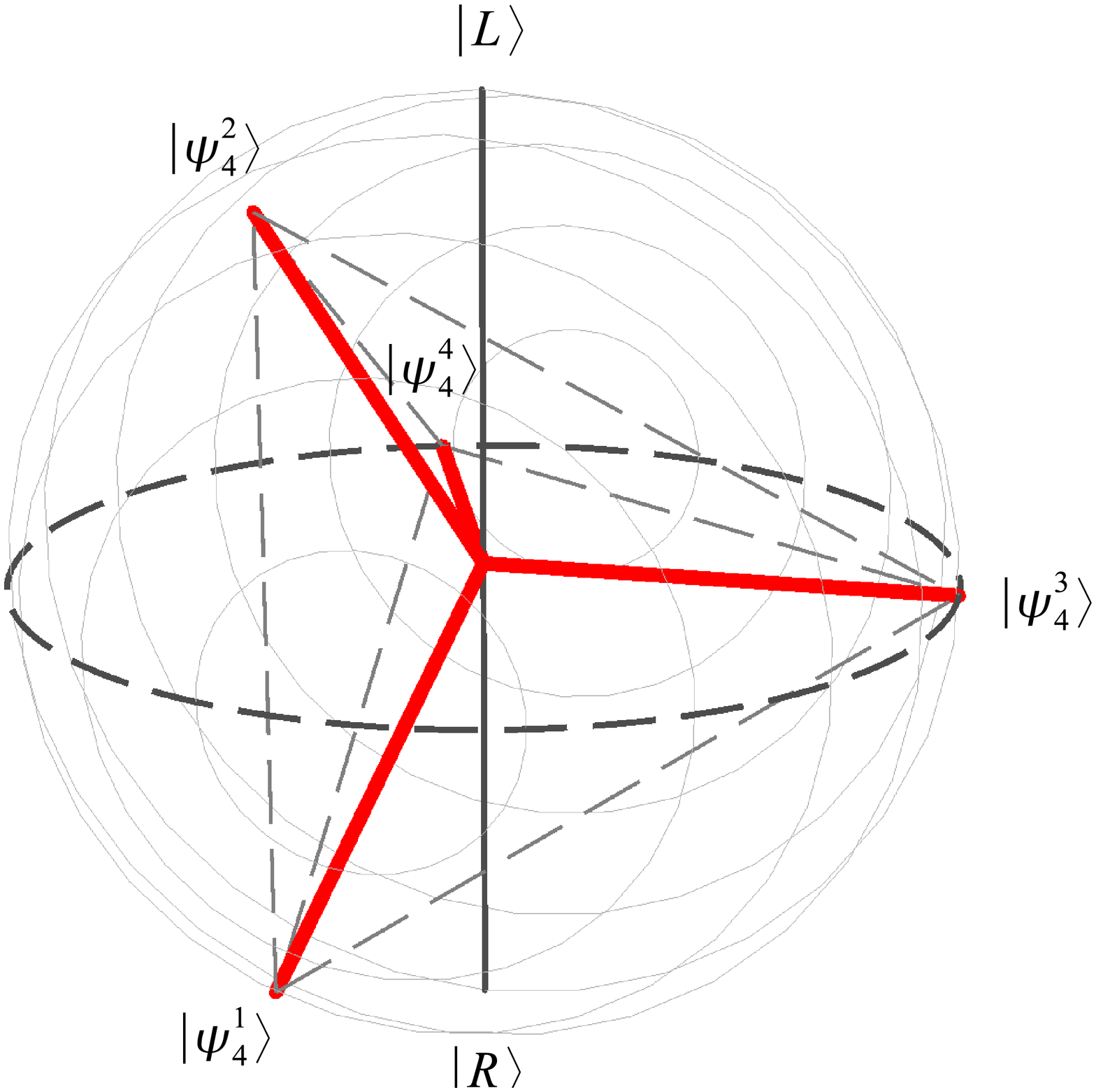}
\end{center}
\caption{Representation of the trine (left) and tetrad (right) states on the Poincar\'e sphere.}
\label{symstates}
\end{figure}

 Finding a minimum error strategy for discriminating between more than two states is, in general
 a difficult problem, although very general statements about the solution can be made for
 qubits \cite{Hunterthesis}.  For the trine ensemble of three equiprobable linear polarisation states
 \begin{eqnarray}
\ket{\psi^3_1} &=& \ket{H}  \nonumber \\
\ket{\psi^3_2} &=& - \frac{1}{2} \ket{H} + \frac{\sqrt{3}}{2} \ket{V} \label{trine} \\
\ket{\psi^3_3} &=& - \frac{1}{2} \ket{H} - \frac{\sqrt{3}}{2} \ket{V} \nonumber
\end{eqnarray}
and the tetrad ensemble of four equiprobable states
\begin{eqnarray}
\ket{\psi^4_1} &=& \frac{1}{\sqrt 3}\left(-\ket{H} + \sqrt{2}e^{-i2\pi i/3}\ket{V}\right) \nonumber \\
\ket{\psi^4_2} &=& \frac{1}{\sqrt 3}\left(-\ket{H} + \sqrt{2}e^{i2\pi i/3}\ket{V}\right) \nonumber \\
\ket{\psi^4_3} &=& \frac{1}{\sqrt 3}\left(-\ket{H} + \sqrt{2}\ket{V}\right) \nonumber \\
\ket{\psi^4_4} &=& \ket{H} 
\end{eqnarray}
the square root measurement is readily shown to give the minimum probability for error.
The trine states are states of linear polarisation separated by $60^\circ$ and the tetrad states
are two states of linear polarisation and two of elliptical polarisation.  In each case they form 
a set of maximally separated points on the surface of the Poincar\'e sphere, as shown in Fig. \ref{symstates}.

\begin{figure}[h]
\begin{center}
\includegraphics[width=80truemm]{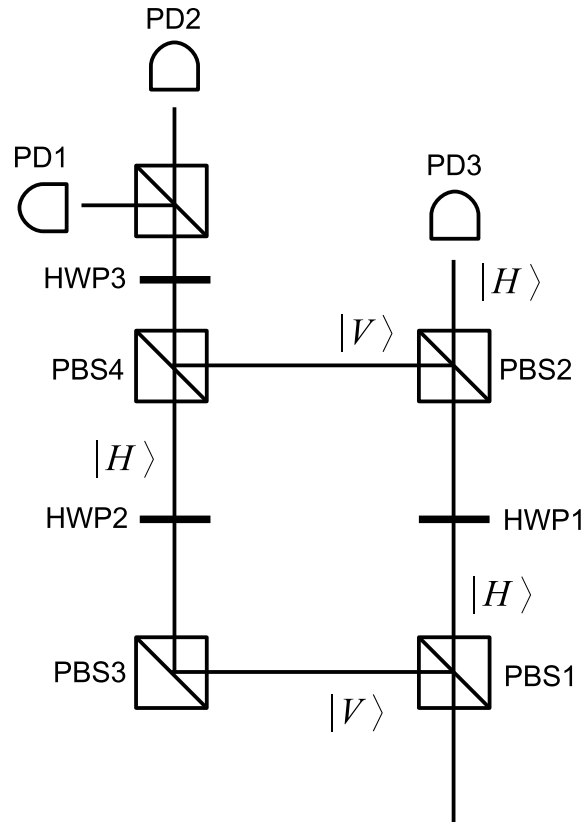}
\end{center}
\caption{Schematic of the Clarke \emph{et al.} experimental realisation of minimum error discrimination between the trine states.  PBS1-3 = polarising beam splitters, HWP1-3 = half waveplates, PD1-3 = photodetectors.  For details see \cite{ClarkeTrine}.}
\label{trineexpt}
\end{figure}

In order to measure more than two orthogonal states of polarisation we need to introduce
an additional degree of freedom and a suitable one is provided by the path of the light beam. 
We shall illustrate this idea only for the trine ensemble, the experimental set-up for which is shown in Fig. \ref{trineexpt}.  Details for the tetrad ensemble can
be found in \cite{ClarkeTrine}.
The input polarising beam splitter separates, coherently, the polarisation 
components by transmitting the horizontal component and reflecting the vertical component.
This allows us to manipulate these components independently.  A half-wave plate placed in 
the path of the horizontally polarised beam rotates the polarisation so that only the requisite
fraction of it is transmitted at the next polarising beam splitter.  The vertically polarised beam is
transformed into a horizontally polarised beam so that it can be recombined coherently with what is 
left of the originally horizontally polarised beam.  Thus the polarisation of this combined
beam is analysed using a final polarising beam splitter.  The photon ends up in one of the three
photodetectors and we can think of each of the trine polarisation states being
transformed into a superposition of exit paths from the interfermometer \cite{ClarkeTrine}:
\begin{eqnarray}
\ket{\psi^3_1} &\rightarrow & -\frac{1}{\sqrt{6}}\ket{P3} - \frac{\sqrt{2}}{\sqrt{3}}\ket{P1}
-\frac{1}{\sqrt{6}}\ket{P3} \nonumber \\
\ket{\psi^3_2} &\rightarrow & -\frac{1}{\sqrt{6}}\ket{P3} + \frac{1}{\sqrt{6}}\ket{P1}
+\frac{\sqrt{2}}{\sqrt{3}}\ket{P3} \nonumber \\
\ket{\psi^3_3} &\rightarrow & \frac{\sqrt{2}}{\sqrt{3}}\ket{P3} +\frac{1}{\sqrt{6}} \ket{P1}
-\frac{1}{\sqrt{6}}\ket{P3} ,
\end{eqnarray} 
where a photon in path $Pi$ will be detected in photodetector $i$.  This measurement device
is optimal as it correctly identifies the initial polarisation state with probability $\frac{2}{3}$.

\subsection{Unambiguous Discrimination}

\begin{figure}[h]
\begin{center}
\includegraphics[width=120truemm]{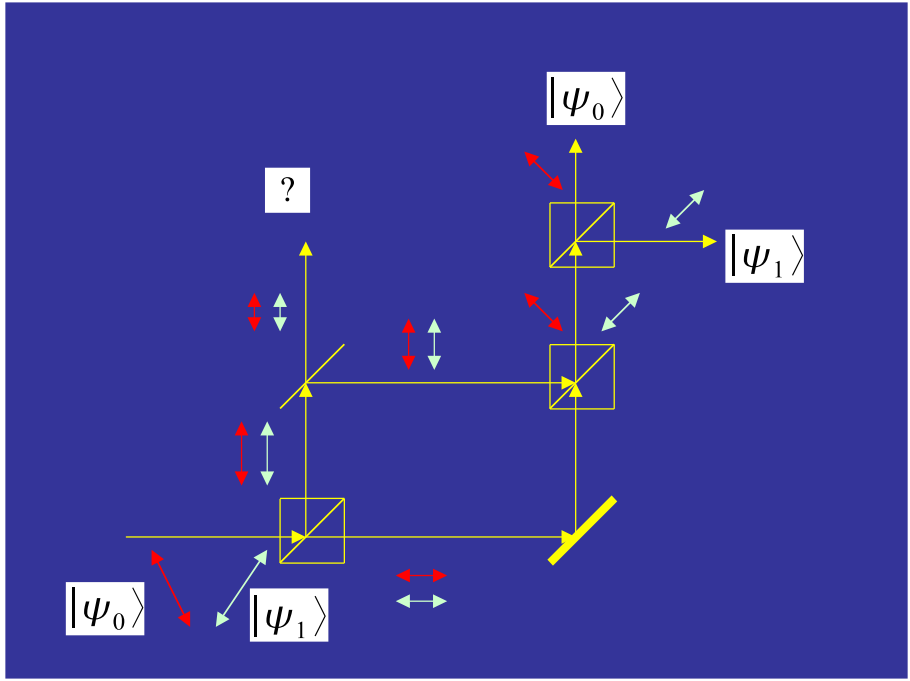}
\end{center}
\caption{Schematic of the Clarke \emph{et al.} experimental realisation of unambiguous discrimination between two non-orthogonal polarisation states.}
\label{unambexpt}
\end{figure}

Unambiguous discrimination between non-orthogonal polarisation states, like the minimum
error measurements described above, requires an extension of the two-dimensional state space
and an interferometer is the ideal device for implementing this.  The idea is depicted in Fig. \ref{unambexpt}.
We have two possible linear polarisation states, each of which has a larger vertical component
of polarisation than horizontal.  The double-headed arrows are intended to represent
the magnitudes of the probability amplitudes at various places.  The input polarising beam 
splitter reflects the vertical component and transmits the horizontal component.  The mirror in
the upper arm of the interferometer transmits just enough for the reflected field to have the 
same amplitude as that in the lower arm.  If the photon escapes from the interferometer at this
point then the measurement is inconclusive.  If it does not, however, then the amplitudes for
the vertical and horizontal fields are equal in magnitude and become othogonal when recombined 
at the output polarising beam splitter.  At this stage they can be discriminated with certainty 
using a final, suitably oriented, polarising beam splitter.  

\begin{figure}[h]
\begin{center}
\includegraphics[width=120truemm]{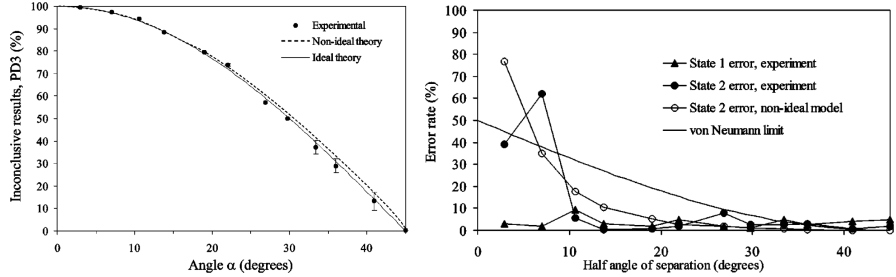}
\end{center}
\caption{Results of the Clarke \emph{et al.} experimental realisation of unambiguous discrimination between two non-orthogonal polarisation states.  The rate of inconclusive results is shown on the left, and the error rate for each initial state given on the right.  A model taking into account the non-ideal characteristics of the beamsplitters was used to generate the non-ideal theory plots in each case.  For full details see \cite{ClarkeUnam}. Copyright (2001) by the Americal Physical Society.}
\label{unambdata}
\end{figure}

The first demonstration of unambiguous discrimination between non-orthogonal polarisation
states used a specially selected length of polarisation maintaining fibre \cite{Huttner96}.  This
has the effect of maintaining, with low losses, the horizontal component of polarisation but
attenuating the orthogonal vertical component.  If the length of the fibre is chosen appropriately then 
any light exiting the fibre will be in one of two orthogonal polarisations and so can be 
discriminated with certainty.  An interferometric experiment has the advantage that it allows us
to measure also the ambiguous results.  The experimental setup \cite{ClarkeUnam} is very similar 
to that for the minimum error discrimination of the three trine states, but with the three measured
outputs now corresponding to the unambiguous identification of the states $\ket{\psi_0}$,
$\ket{\psi_1}$ and to the ambiguous result.  The results of this experiment are shown in Fig. \ref{unambdata}.

We have presented here only the simplest experiments, but more complicated problems have
also been addressed.  In particular, unambiguous discrimination has been demonstrated for
three possible states and also between non-orthogonal pure and mixed states \cite{Mohseni04}.
The generalised measurements described here have all been implemented using light but
the principles are independent of the system used.  It should be noted, particularly in the context
of quantum information, that non-orthogonal states encoded in the energy levels of atoms or ions
can similarly be subjected to generalised measurements with unoccupied levels used to assist
in the process \cite{Franke-Arnold01}.

\subsection{Maximum confidence measurements}
\begin{figure}[t]
\begin{center}
\includegraphics[width=90truemm]{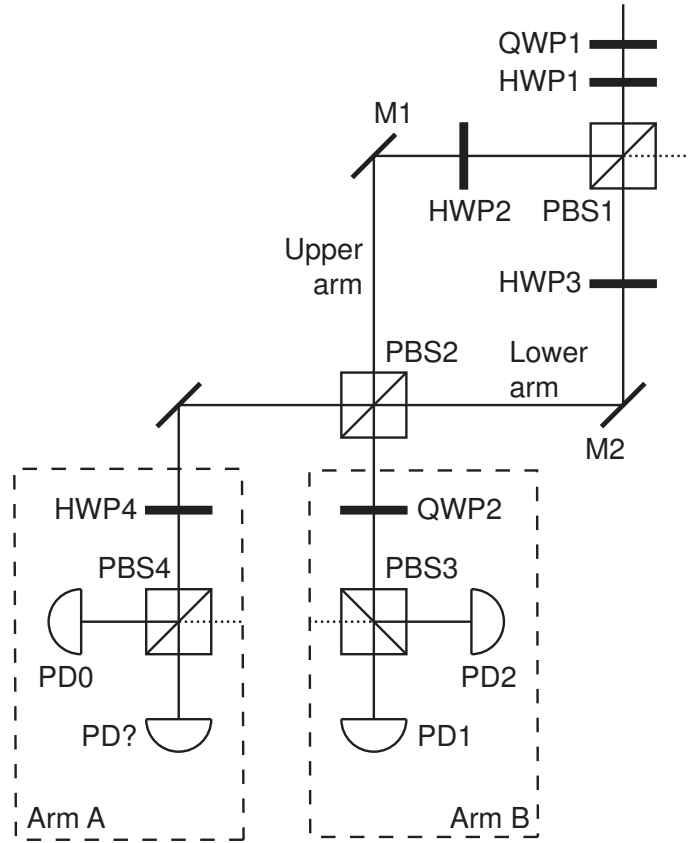}
\end{center}
\caption{Schematic of the experimental apparatus used to demonstrate maximum confidence discrimination between three elliptical polarisation states. PBS1-4 = polarising beamsplitters, QWP1-2 = quarter waveplates, HWP 1-4 = half waveplates, PD0-2, PD? = photodetectors.}
\label{mcexpt}
\end{figure}
Maximum confidence discrimination between three symmetric states in two dimensions (the simplest possible case), has also been demonstrated experimentally using the polarisation of light as a qubit \cite{Mosley06,Croke07}.  In the experimental realisation, the states given in equation (\ref{states}) were encoded in the left/right circular polarisation basis, and the set-up distinguished between the elliptical polarisations
\begin{eqnarray}
\ket{\psi_0} &=& \cos \theta \ket{R} + \sin \theta \ket{L}, \nonumber \\
\ket{\psi_1} &=& \cos \theta \ket{R} + e^{2 \pi i/3} \sin \theta \ket{L}, \nonumber \\
\ket{\psi_2} &=& \cos \theta \ket{R} + e^{-2 \pi i/3} \sin \theta \ket{L}.
\end{eqnarray}
The maximum confidence measurement for these states, as we have seen, has four outcomes, one corresponding to each possible state and one inconclusive result.  The apparatus used is shown in Fig. \ref{mcexpt}, and again features an interferometer to provide the extension to the state space necessary to realise all four outcomes.  In this set-up, the outcomes 0 and ? are grouped together in one output arm of the interferometer, while the other arm corresponds to outcomes 1 and 2.  Thus two detectors placed in the output arms A and B of the apparatus would realise the two outcome generalised measurement described by the POM $\{ \hat{\pi}_? + \hat{\pi}_0, \hat{\pi}_1 + \hat{\pi}_2\}$.  In fact this set-up is completely general, and by appropriate choice of orientations of the waveplates QWP1 and HWP1-3, may be used to implement any such two-element measurement.  Further, any $N$ outcome measurement may be, in principle, performed using a number of such modules in series \cite{Croke07,Ahnert05}.  Thus, after PBS2, two orthogonal modes in arm A correspond to outcomes 0 and ?, while two orthogonal modes in arm B correspond to results 1 and 2.  Finally HWP4, QWP2 and PBS3-4 are used to separate these modes, which are then detected at the photodetectors in the output arms.
\begin{figure}[h]
\begin{center}
\includegraphics[width=90truemm]{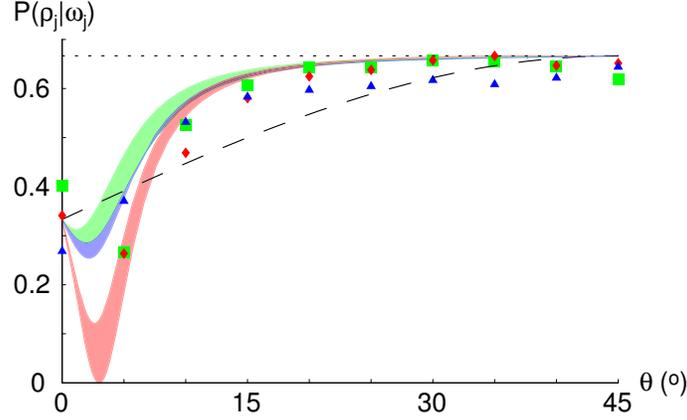}
\end{center}
\caption{Results of the maximum confidence discrimination experiment.  Graph shows the confidence figure of merit for measurement outcomes 0 (red), 1 (green) and 2 (blue).  Lines indicate the theoretical value of the figure of merit for the maximum confidence (dotted) and minimum error (dashed) measurement strategies.  Shaded areas indicate the range of values consistent with a non-ideal model, taking into account errors introduced at the polarising beamsplitters, for details see \cite{Croke07}.  Figure reproduced from \cite{Mosley06}, copyright by the American Physical Society.}
\label{mcdata}
\end{figure}
The results of this experiment demonstrated an improvement over the minimum error measurement in the confidence figure of merit for linearly dependent states and are shown in Fig. (\ref{mcdata}).

\subsection{Mutual information}

The strategies for maximising the mutual information for two pure states require us to perform a
minimum error measurement \cite{Levitin}.  With more states we require, in general, a generalised
measurement \cite{Davies,Sasaki99}.  For the trine and tetrad states we obtain the accessible 
information by eliminating, with certainty one of the possible states.  This can be realised experimentally
using the same device as that devised for the minimum error measurement, simply by interchanging 
everywhere the horizontal and vertical components of polarisation.  In other words, the device
for maximising the mutual information for the trine or tetrad states is the same as that for
minimising the error in discriminating between a set of states \emph{orthogonal} to the given
trine or tetrad.  For more than four states of linear polarisation, we can maximise the mutual information by performing a measurement with just three possible outcomes \cite{Sasaki99}.

The experiment to realise the minimum error discrimination between two non-orthogonal
polarisation states \cite{Riis} also provided the maximum mutual information.  For the 
pure states (\ref{Twostates}) with $\theta = 15^\circ$, corresponding to linear polarisations 
at an angle of $30^\circ$, the mutual information derived from the measurements was
\cite{Barnett04}
\begin{equation}
H^{\rm 2 states}(A:B) = 0.196 \pm 0.007 \: \: {\rm bits},
\end{equation}
which compares well with the theoretical value of $0.189$ bits.  For the trine and the tetrad 
\cite{ClarkeTrine} we found
\begin{eqnarray}
H^{\rm trine}(A:B) &=& 0.491^{+0.011}_{-0.027}\: \:  {\rm bits} \nonumber \\
H^{\rm tetrad}(A:B) &=& 0.363^{+0.09}_{-0.024} \: \: {\rm bits} ,
\end{eqnarray}
which should be compared with the theoretical values of $0.585$ bits and $0.415$ bits respectively.
It is important to note that these experimental values are good enough to demonstrate the
necessity of performing a generalised measurement as the theoretical maximum mutual 
information for the trine and tetrad states using conventional projective measurements are
$0.459$ bits and $0.311$ bits respectively.  A subsequent more careful experiment 
produced a substantially higher value for the mutual information obtained using the trine
ensemble of $0.556$ bits and also realised the optimal measurements for sets of five and 
seven states of linear polarisation \cite{Mizuno}.

\section{Conclusion}

Quantum theory allows us to prepare, at least in principle, even the simplest system in an uncountable
infinity of different ways.  The polarisation for a single photon, for example, can be prepared in
a state that correcsponds to any point on the surface of the Poincar\'e sphere.  It is a fundamental
consequence of the superposition principle, however, that no measurement can discriminate 
with certainty between two non-orthogonal quantum states.  The challenge for quantum state
 discrimination is to perform this task as well as is possible.  

It is evident that selecting the best possible measurement in any given situation usually requires
us to perform a generalised measurement.  These are general in the sense that they represent,
not just projective measurements of the kind envisaged by von Neumann \cite{vonNeumann},
but rather the most general measurements possible within the confines of quantum theory.
The POM formalism is, as we have seen, a remarkable tool in the search for optimal 
measurements.  That this is the case is a consequence of the facts that (i) any set of
probability operators satisfying the required properties listed in section 2 correspond to
a possible quantum measurement and (ii) all possible measurements can be described by
an appropriate set of probability operators.  This means that we can separate the mathematical
task of finding the theoretically optimum measurement from the practical one of designing a
measurement to implement it.

We have seen that optimal measurements have been found to minimise the error in identifying
the state, discriminate between states unambiguously and to determine the state with the maximum 
level of confidence.  These similar sounding goals are all subtly different and correspond, for the
most part, to quite distinct measurements.  We have also discussed yet another task relevant
to quantum communications, that of maximising the information transferred.  The problem of state
discrimination acquired much of its significance from considering the problem of quantum
communication and in particular from quantum cryptography 
\cite{Phoenix95,Lo,Bouwmeester,Gisin02,Loepp06,Barnett09}.  All existing implementations
of these are based on optics and it is perhaps not surprising, therefore, that it is in optics 
that the experimental advances in quantum state discrimination have been made.  We have
described, in particular, how quantum limited measurements have been devised on optical
polarisation to realise the optimal measurements for detection with minimum error, unambiguous
discrimination  as well as detection with maximum confidence and maximum mutual information.
As quantum information technology develops the ability to optimise performance by performing
the best possible measurements can only become more important.

\vskip1cm

\textbf{Acknowledgements}

\vskip0.5cm

This work was supported, in part, by the UK Engineering and Physical Sciences Research Council,
the Royal Society and the Wolfson Foundation (SMB), by the Synergy fund of the Universities of Glasgow and Strathclyde, and by Perimeter Institute for Theoretical Physics (SC). Research at Perimeter Institute is supported by the Government of Canada through Industry Canada and
by the Province of Ontario through the Ministry of Research \& Innovation.

\end{document}